# Power-Domain Non-Orthogonal Multiple Access (NOMA) in 5G Systems: Potentials and Challenges

S.M. Riazul Islam, Nurilla Avazov, Octavia A. Dobre, and Kyung-Sup Kwak

*Abstract*—**Non-orthogonal multiple access (NOMA) is one of the promising radio access techniques for performance enhancement in next-generation cellular communications. Compared to orthogonal frequency division multiple access (OFDMA), which is a well-known high-capacity orthogonal multiple access (OMA) technique, NOMA offers a set of desirable benefits, including greater spectrum efficiency. There are different types of NOMA techniques, including power-domain and code-domain. This paper primarily focuses on power-domain NOMA that utilizes superposition coding (SC) at the transmitter and successive interference cancellation (SIC) at the receiver. Various researchers have demonstrated that NOMA can be used effectively to meet both network-level and user-experienced data rate requirements of fifth-generation (5G) technologies. From that perspective, this paper comprehensively surveys the recent progress of NOMA in 5G systems, reviewing the state-of-the-art capacity analysis, power allocation strategies, user fairness, and user-pairing schemes in NOMA. In addition, this paper discusses how NOMA performs when it is integrated with various proven wireless communications techniques, such as cooperative communications, multiple-input multiple-output (MIMO), beamforming, space–time coding, and network coding, among others. Furthermore, this paper discusses several important issues on NOMA implementation and provides some avenues for future research.**

*Index Terms*—**Non-Orthogonal Multiple Access (NOMA), Orthogonal Multiple Access (OMA), 5G, NOMA Solutions, NOMA Performance, Research Challenges, Implementation Issues.**

## I. INTRODUCTION

From analog phone calls through to all Internet Protocol services, including voice and messaging, each transition has been encouraged by the need to meet the requirements of the new generation of mobile technology. Subsequently, mobile communications technology is presently

This work was supported by National Research Foundation of Korea-Grant funded by the Korean Government (Ministry of Science, ICT and Future Planning)-NRF-2014K1A3A1A20034987), as well as the Natural Sciences and Engineering Research Council of Canada (NSERC) through its Discovery Program.

S. M. R. Islam (e-mail: islam.smriaz@gmail.com), N. Avazov (e-mail: nurilla.avazov@inha.ac.kr), K. S. Kwak (e-mail: kskwak@inha.ac.kr), are with the UWB Wireless Communications Research Center, Inha University, South Korea.

O. A. Dobre (e-mail: odobre@mun.ca) is with the Faculty of Engineering and Applied Science, Memorial University, Canada.

facing a new challenge, giving birth to a hyper-connected society through the emergence of fifth-generation (5G) services. With enormous potential for both consumers and industry, 5G is expected to roll out by 2020. From the next-generation radio access technology viewpoint, a step change in data speed and a significant reduction in end-to-end latency is a major concern for 5G, since the rapid development of the mobile Internet and the Internet of Things (IoT) exponentially accelerates the demand for high data–rate applications. In particular, many of the industry initiatives that have progressed with work on 5G declare that the network-level data rate in 5G should be 10-20 Gbps (that is, 10-20 times the peak data rate in 4G), and the user-experienced data rate should be 1 Gbps (100 times the user-experienced data rate in 4G). They also set the latency (end-to-end round-trip delay) at 1 millisecond (one-fifth of the latency in 4G).

The underlying physical connection in a cellular network is called radio access technology, which is implemented by a radio access network (RAN). A RAN basically utilizes a channel access technique to provide the mobile terminals with a connection to the core network. The design of a suitable multiple access technique is one of the most important aspects in improving the system capacity. Multiple access techniques can broadly be categorized into two different approaches [1], namely, orthogonal multiple access (OMA) and non-orthogonal multiple access (NOMA). An orthogonal scheme allows a perfect receiver to entirely separate unwanted signals from the desired signal using different basis functions. In other words, signals from different users are orthogonal to each other in orthogonal schemes. Time division multiple access (TDMA), and orthogonal frequency-division multiple access (OFDMA) are a couple of examples of OMA schemes. In TDMA, several users share the same frequency channel on a time-sharing basis. The users communicate in rapid succession, one after the other, each using their assigned time slots. OFDMA allows multi-user communications through an orthogonal frequency-division multiplexing (OFDM) technique in which subcarrier frequencies are chosen so that the subcarriers are orthogonal to each other. In contrast to OMA, NOMA allows allocating one frequency channel to multiple users at the same time within the same cell and offers a number of advantages, including improved spectral efficiency (SE), higher cell-edge throughput, relaxed channel feedback (only the received signal strength, not exact channel state information (CSI), is required), and low transmission latency (no scheduling request from users to base station is required). The available NOMA techniques can broadly be divided into two categories, namely, power-domain and code-domain NOMA. This paper focuses on the power-domain NOMA that superposes multiple users in power domain and exploits the channel gain difference between multiplexed users. At the transmitter side, signals from various users are superposed and the resulting signal is then transmitted over the same channels (i.e., the same time-frequency resources). At the receiver sides, multiuser detection (MUD) algorithms, such as successive interference cancellation (SIC) are utilized to detect the desired signals. Although this paper primarily surveys the power-domain superposition coding (SC)-based NOMA, a brief discussion of other classes of NOMA is given in the next section.

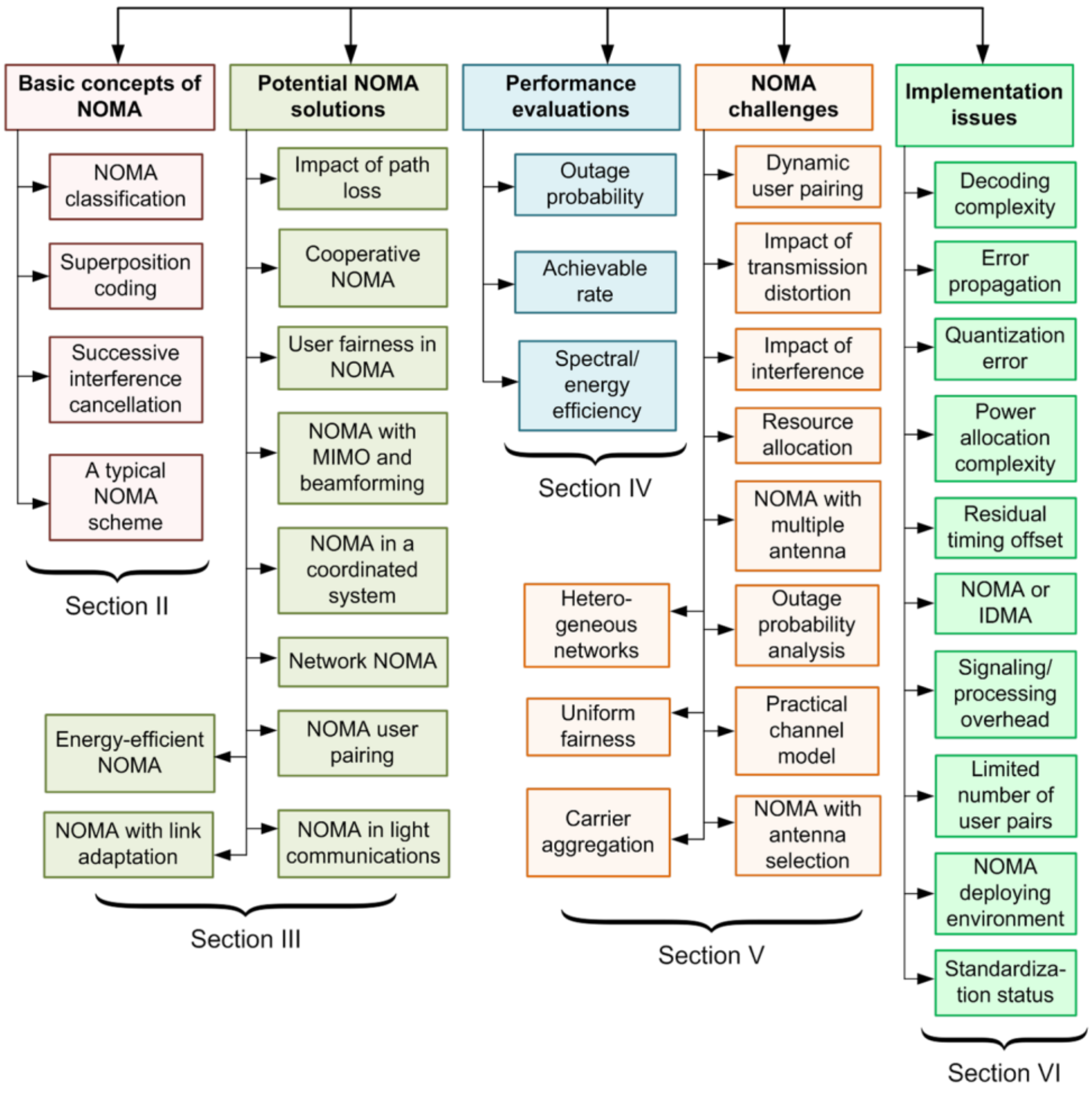

Fig. 1. The organization and structure of this paper.

OMA is a realistic choice for achieving good performance in terms of system-level throughput. However, due to the aforementioned upcoming wave, 5G networks require further enhancement in system efficacy. In this regard, researchers all over the globe have started investigating NOMA as a promising multiple access scheme for future radio access. NOMA achieves superior spectral efficiencies by SC at the transmitter with SIC at the receiver [2, 3]. On the top of that, the evolution of wireless networks to 5G poses new challenges for energy efficiency (EE), since the entire network will be ultra-dense. With an extreme increase in the number of infrastructure nodes, total energy consumption may simply surpass an acceptable level. Although substantial energy is basically consumed by the hardware, NOMA has an inherent ability to adapt the transmission strategy according to the traffic and users' CSIs. Thus, it can achieve a good operating point, where both spectrum efficiency and EE become optimum.

Over the past few years, NOMA has attracted a great deal of attention from researchers trying to meet 5G requirements. As a consequence, many research efforts in this field already exist. Research trends in NOMA include diverse topics, for example, various performance analysis methods, fairness analysis, EE, and user pairing. Many researchers are attempting to further enhance the performance of other existing wireless technologies, such as cooperative communications, multiple-input multiple-output (MIMO), light communications, and relay networks by using NOMA. However, NOMA in 5G is still in its infancy. At this stage, comprehensive knowledge on the up-to-date research status of NOMA in 5G systems is extremely useful to researchers who want to do more in this area. In this regard, this paper examines the trends in NOMA-based research and the various issues that need to be addressed to transform radio access techniques through NOMA innovation. The rest of this paper is organized as follows. In Section II, the basic concepts of NOMA are presented, whereas Section III discusses state-of-the-art NOMA research into solutions. Section IV evaluates the performance of various NOMA solutions. In Section V, several challenges to NOMA implementations are presented, and Section VI discusses other implementation issues. The final section concludes this survey. Fig. 1 presents an overview of the organization and structure of this paper.

## II. Basic Concepts Of NOMA

There exist different NOMA solutions, which can primarily be classified into two major approaches. Fig. 2 presents a simple classification of the existing NOMA techniques. Unlike power-domain NOMA, which attains multiplexing in power domain, code-domain NOMA achieves multiplexing in code domain. Like the basic code division multiple access (CDMA) systems, code-domain NOMA shares the entire available resources (time/frequency). In contrast, code-domain NOMA utilizes user-specific spreading sequences that are either sparse sequences or non-orthogonal cross-correlation sequences of low correlation coefficient. This can be further divided into a few different classes, such as low-density spreading CDMA (LDS-CDMA) [4, 5], low-density spreading-based OFDM (LDS-OFDM) [6, 7], and sparse code multiple access (SCMA) [8, 9]. The use of low-density spreading sequences helps LDS-CDMA to limit the impact of interference on each chip of basic CDMA systems. LDS-OFDM can be thought of as an amalgamation of LDS-CDMA and OFDM, where the information symbols are first spread across low-density spreading sequences and the resultant chips are then transmitted on a set of subcarriers. SCMA is a recent code-domain NOMA technique based on LDS-CDMA. In contrast to LDS-CDMA, the information bits can be directly mapped to different sparse codewords, because both bit mapping and bit spreading are combined. When compared to LDS-CDMA,

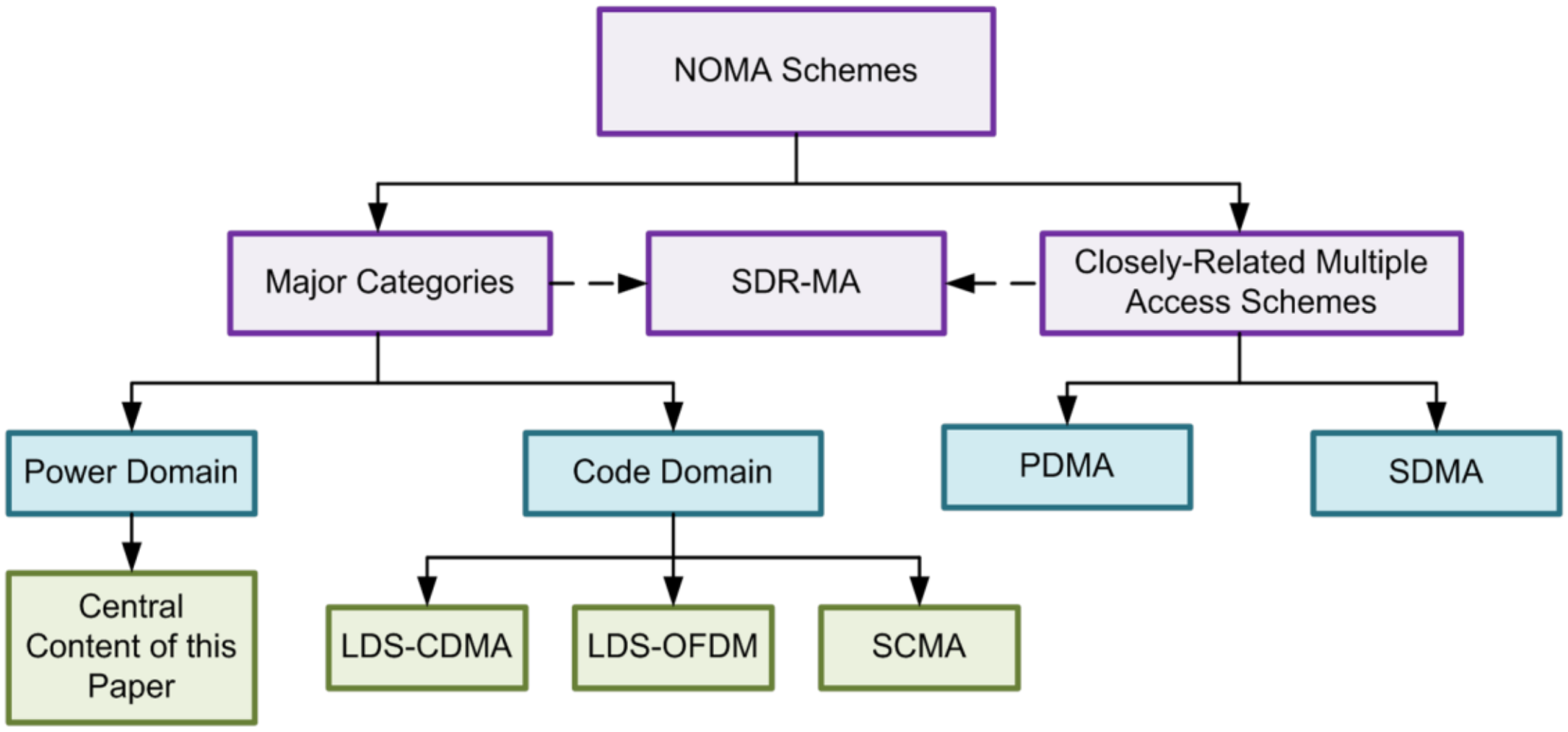

Fig. 2. A simple classification of NOMA techniques.

SCMA provides a low complexity reception technique and offers improved performances.

There exist some other multiple access techniques, which are also closely-related to NOMA, including pattern division multiple access (PDMA) [10] and spatial division multiple access (SDMA) [11, 12, 13, 14]. PDMA can be realized in various domains. At the transmitter side, PDMA first maximizes the diversity and minimizes the overlaps among multiple users in order to design non-orthogonal patterns. The multiplexing is then performed either in the code domain, spatial domain, or a combination of them. For SDMA, the working principle is inspired by basic CDMA systems. Instead of using user-specific spreading sequences, SDMA distinguishes different users by using user-specific channel impulse responses (CIRs). This technique is particularly useful for the cases where the number of uplink users is considerably higher than the number of corresponding receiving antennas in BS. However, accurate CIR estimation becomes challenging for a large number of users. The concept of software defined radio for multiple access (SDR-MA) allows various forms of NOMA schemes to coexist [15]. This technique provides a flexible configuration of participating multiple access schemes in order to support heterogeneous services and applications in 5G. It is worth noting that while the aforementioned list provides some insights into different forms of NOMA, it is not exhaustive, and the primary focus of this paper is on power-domain NOMA.

In the following, a brief note about SC and SIC is presented, since these two basic techniques play important roles in understanding the class of NOMA on which this paper focuses on. Henceforth, this paper refers to power-domain NOMA simply by NOMA.

### *A. Superposition Coding (SC)*

The SC that was first proposed in [16] is a technique of simultaneously communicating information to several receivers by a single source. In other words, it allows the transmitter to transmit multiple users' information at the same time. Examples of communications in a superposed fashion include broadcasting a television signal to multiple receivers and giving a speech to a group of people with different backgrounds and aptitudes, such as a lecture in a classroom. Suppose a professor is giving students information through a lecture in the classroom. Since there are differences among the qualities and backgrounds of the students, some of them receive most of the information, and others receive only a little. The lecture may be organized that it proceeds at the pace of the student with the weakest background. However, in the ideal situation, the lecture can be designed in such a way that the students with the most suitable background obtain more information and the poor students get at least the minimum amount of information. This situation is an example of a broadcast channel where a superposed lecture is being delivered by the speaker. To make SC practical, the transmitter must encode information relevant to each user. For example, for a two-user scenario, the transmitter will have to contain two point-to-point encoders that map their respective inputs to complex-valued sequences of the two-user signal. In order to show how SC is performed, a schematic diagram is given in Fig. 3, where the quadrature phase-shift keying (QPSK) constellation of user 1 with higher transmitting power is superposed on that of user 2 with lower transmitting power. It can be mentioned that SC is a recognized non-orthogonal scheme that attains the capacity on a scalar Gaussian broadcast channel. Vanka et al. introduced good strategies for SC [17], and proposed a design technique for SC by using off-the-shelf single-user coding and decoding blocks. In the superposition encoding phase, two point-to-point encoders, $f_1: \{0,1\}^{\lfloor 2^{TR_1} \rfloor} \rightarrow C^T$ and $f_2: \{0,1\}^{\lfloor 2^{TR_2} \rfloor} \rightarrow C^T$ first map the respective input bits to two output bit sequences $S_1(n)$ and $S_2(n)$, respectively, each of block length $T$. Here $R_1$ and $R_2$ denote the transmission rates of user 1 and user 2, respectively, and $\lfloor \cdot \rfloor$ represents the floor operator. $C$ is nothing but a code library. Then, a summation device provides an output sequence as

$$X(n) = \sqrt{P\beta_1} S_1(n) + \sqrt{P\beta_2} S_2(n), \tag{1}$$

where $\beta_i$ represents a fraction of the total power $P$ assigned to user $i$, subject to the constraint on $\beta_1 + \beta_2 = 1$.

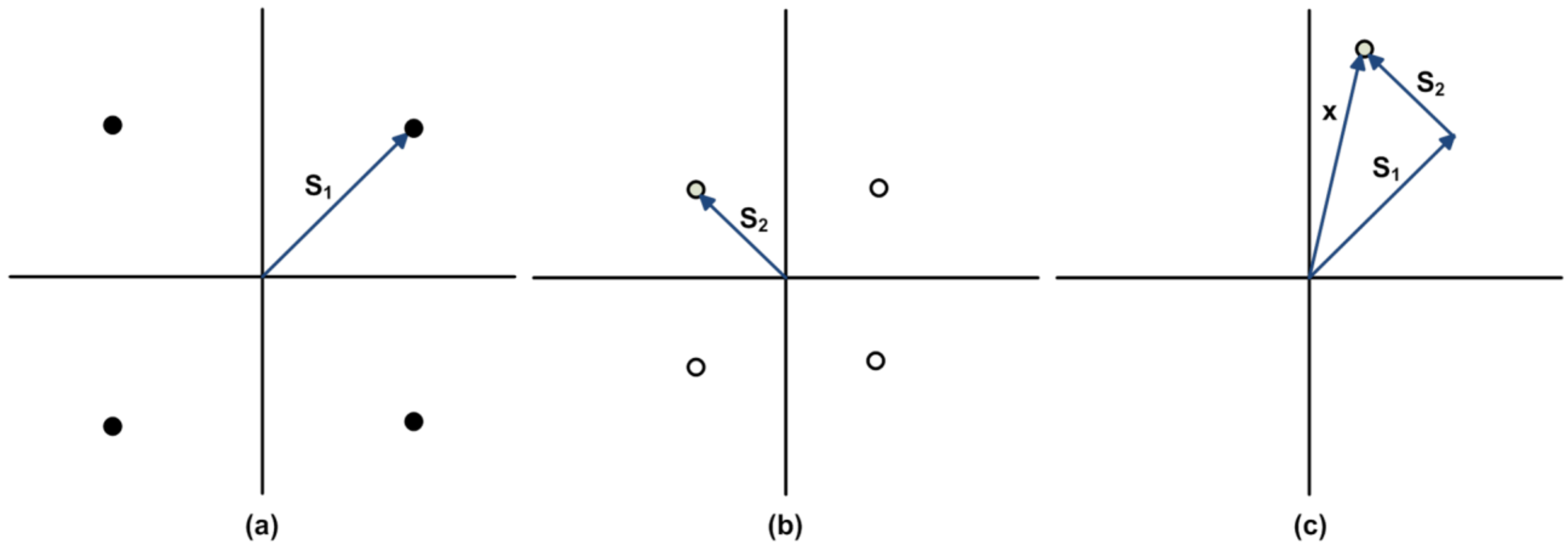

Fig. 3. An example of SC encoding (a) signal constellation of user 1 (b) signal constellation of user 2 (c) constellation of superposed signal.

*B. Successive Interference Cancellation (SIC)*

To decode the superposed information at each receiver, Cover first proposed the SIC technique [16]. SIC is conceivable by exploiting specifications on the differences in signal strength among the signals of interest. The basic idea of SIC is that user signals are successively decoded. After one user's signal is decoded, it is subtracted from the combined signal before the next user's signal is decoded. When SIC is applied, one of the user signals is decoded, treating the other user signal as an interferer, but the latter is then decoded with the benefit of the signal of the former having already been removed. However, prior to SIC, users are ordered according to their signal strengths, so that the receiver can decode the stronger signal first, subtract it from the combined signal, and isolate the weaker one from the residue. Note that each user is decoded treating the other interfering users as noise in signal reception. Fig. 4 presents the technique for decoding the superposed signal (Fig. 3) at the receiving side. Here, the constellation point of user 1 is decoded first from the received signal. Then, the decoding of the constellation point of user 2 is performed with respect to decoded constellation point of user 1. To gain a deeper understanding of how SIC performs in wireless communications in general, and in OFDM (and MIMO systems) in particular, interested readers are referred to [18]. In brief, the particular process involved in decoding the superposed messages can be mathematically expressed as follows [17]:

1) At user 1, a single-user decoder $\mathrm{g}_1 : C^T \to \{0,1\}^{2^{TR_1}}$ decodes the message $S_1(n)$ by treating $S_2(n)$ as noise.
2) User 2 performs the following steps to successively recover its message from its received signal $Y_2(n)$:

a) Decode user 1's message $S_1(n)$ by using the single-user decoder $\mathrm{g}_1 : C^T \to \{0,1\}^{2^{TR_1}}$.
b) Subtract $\sqrt{P\beta_1}h_2 S_1(n)$ from the received signal $Y_2(n)$

$$Y_2'(n) = Y_2(n) - \sqrt{P\beta_1}h_2 S_1(n), \tag{2}$$

where $h_2$ is the complex channel gain at user 2.

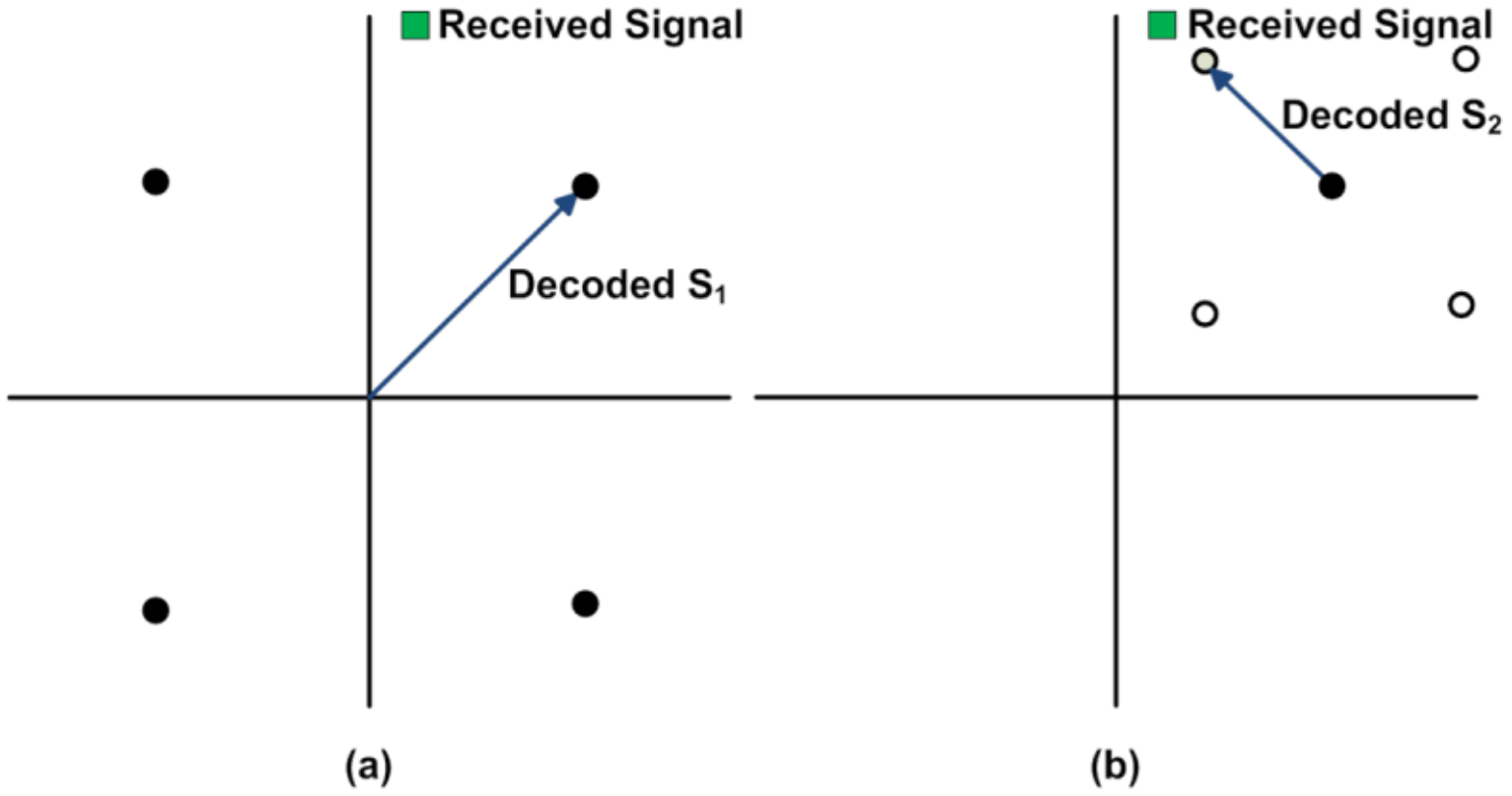

Fig. 4. An example of SC decoding (a) decoding the signal of user 2 (b) decoding the signal of user 1.

c) Decode user 2's messages $S_2(n)$ by applying another single-user decoder $\mathrm{g}_2 : C^T \to \{0,1\}^{2^{TR_2}}$ on $Y_2'(n)$.

*C. A Typical NOMA Scheme*

This scheme considers a single-cell downlink scenario where there is a single base station (BS), and $N$ users $U_i$, with $i \in \mathrm{N} = \{1,2, \dots, N\}$, and all terminals are equipped with a single antenna. Note that a similar uplink scenario can also be described, and a NOMA scheme can equally be utilized there. The BS always sends data to all users simultaneously, subject to the constraint of total power $P$. It is assumed that the wireless links experience independent and identically distributed (i.i.d.) block Rayleigh fading and additive white Gaussian noise (AWGN). The channels are sorted as $0 < |h_1|^2 \le |h_2|^2 \le \cdots \le |h_i|^2 \dots \le |h_N|^2$, which indicates that user $U_i$ always holds the $i$th weakest instantaneous channel. The NOMA scheme allows simultaneous serving of all users by using the entire system bandwidth (BW) to transmit data with the help of SC at the BS and SIC decoding techniques at the users. Here, user multiplexing is executed in the power domain. The BS transmits a linear superposition of $N$ users' data by allocating a fraction $\beta_i$ of the total power to each $U_i$, i.e., the power allocated for the $i$th user is $P_i = \beta_i P$. On the receiving end, each user decodes the signals of the weaker users, i.e., $U_i$ can decode the signals for each $U_m$ with $m < i$. The signals for weaker users are then subtracted from the received signal to decode the signal of user $U_i$, itself treating the signals of the stronger users, $U_m$, with $m > i$, as interference. The received signal at user $U_i$ can be represented as

$$y_i = h_i x + w_i. \tag{3}$$

Here, $x = \sum_{i=1}^{N} \sqrt{P\beta_i} S_i$ is the superposed signal transmitted by the BS, with $S_i$ being the signal for user $U_i$. Also, $w_i$ is the AWGN of user $U_i$ with zero mean and variance $\sigma_n^2$. If signal superposition at the BS, and SIC at $U_i$, are carried out perfectly, the data rate achievable for user $U_i$ for 1 Hz system BW is given by[1]

$$R_i = \log_2\left(1 + \frac{\beta_i P |h_i|^2}{P|h_i|^2 \sum_{k=i+1}^{N} \beta_k + \sigma_n^2}\right). \tag{4}$$

Note that the data rate of user $U_N$ is $R_N = \log_2(1 + \beta_N P |h_N|^2/\sigma_n^2)$, as this user successively decodes and cancels all other users' signals prior to decoding its own signal.

Also note that a strong user experiences a better channel condition, but that does not mean the signal strength is stronger. In fact, a lower transmit power is assigned to a strong user, and a weak user is assigned more power. Thus, the weak user's signal is the strongest one. Therefore, NOMA does not contradict the basic concept of SIC, that decoding of the strongest signal should be performed first.

Fig. 5 represents the aforementioned NOMA scheme with two users. This figure also represents the OMA scheme to disclose the particular advantages of a NOMA scheme over an OMA one. With NOMA, the entire 1 Hz BW is simultaneously used by two users. However, with OMA, user 1 uses α Hz and the remaining $(1 - \alpha)$ Hz is assigned to user 2. In NOMA, user 2 first performs SIC to decode the signal for user 1, since the channel gain of user 2 is higher than that of user 1. The decoded signal is then subtracted from the received signal of user 2. This resultant signal is eventually used to decode the signal for user 2. For user 1, SIC is not executed, and the signal is directly decoded. Thus, the achievable data rate for user 1 and user 2 are given by (5) and (6), respectively.

$$R_1 = \log_2\left(1 + \frac{P_1 |h_1|^2}{P_2 |h_1|^2 + \sigma_n^2}\right), \tag{5}$$

$$R_2 = \log_2(1 + P_2 |h_2|^2/\sigma_n^2). \tag{6}$$

Under OMA, the achievable data rate for user 1 and user 2 are given by (7) and (8), respectively:

[1] The imperfect SIC causes error propagation in subsequent decoding of the signals of NOMA users. Therefore, the effect of the said error propagation on NOMA performances remains a concern. This issue will be briefly discussed in Section VI.

$$R_1 = \alpha \log_2(1 + P_1|h_1|^2/\sigma_n^2), \quad (7)$$

$$R_2 = (1-\alpha)\log_2\left(1 + \frac{P_2|h_2|^2}{\sigma_n^2}\right). \quad (8)$$

It is clear from (5) and (6) that the NOMA scheme controls the throughput of each user by adjusting the power allocation ratio, $P_1/P_2$. Thus, the overall throughput and user fairness are closely related to the power allocation scheme. If an asymmetric channel, where signal-to-noise ratios (SNRs) of the two users are different, is considered, it can be numerically shown that the values of $R_1$ and $R_2$, calculated from (5) and (6), respectively, are considerably much higher than those of $R_1$ and $R_2$ calculated from (7) and (8). This numerical comparison is basically a special case of the multi-user channel capacity analysis [3]. Fig. 6 gives us the idea of a generalized capacity comparison of NOMA and OMA for two users. It shows that the boundary of achievable rate pairs with NOMA is outside the OMA capacity region, in general. Therefore, NOMA is highly effective in terms of system-level throughput when the channels are different for two users. On that, NOMA is considered a promising multiple-access technique for future radio access.

NOMA can equally be applied in an uplink scenario. In the uplink, SIC is performed at the BS (see Fig. 7). For a two-user NOMA, the received signal at the BS is represented as

$$y = h_1\sqrt{P\beta_1}S_1 + h_2\sqrt{P\beta_2}S_2 + w. \quad (9)$$

Here, user $U_i$ transmits the signal $S_i$, with $P_i = \sqrt{P\beta_i}$ being the transmission power. Also, $w$ is the AWGN with zero mean and variance $\sigma_w^2$. With the help of SIC, the BS decodes the signals of $U_1$ and $U_2$ in two stages. In the first stage, the signal of $U_2$ is decoded, treating the signal of $U_1$ as noise. In the next stage, the receiver subtracts the decoded signal $S_2$ from the received signal and then decodes the signal of $U_1$.

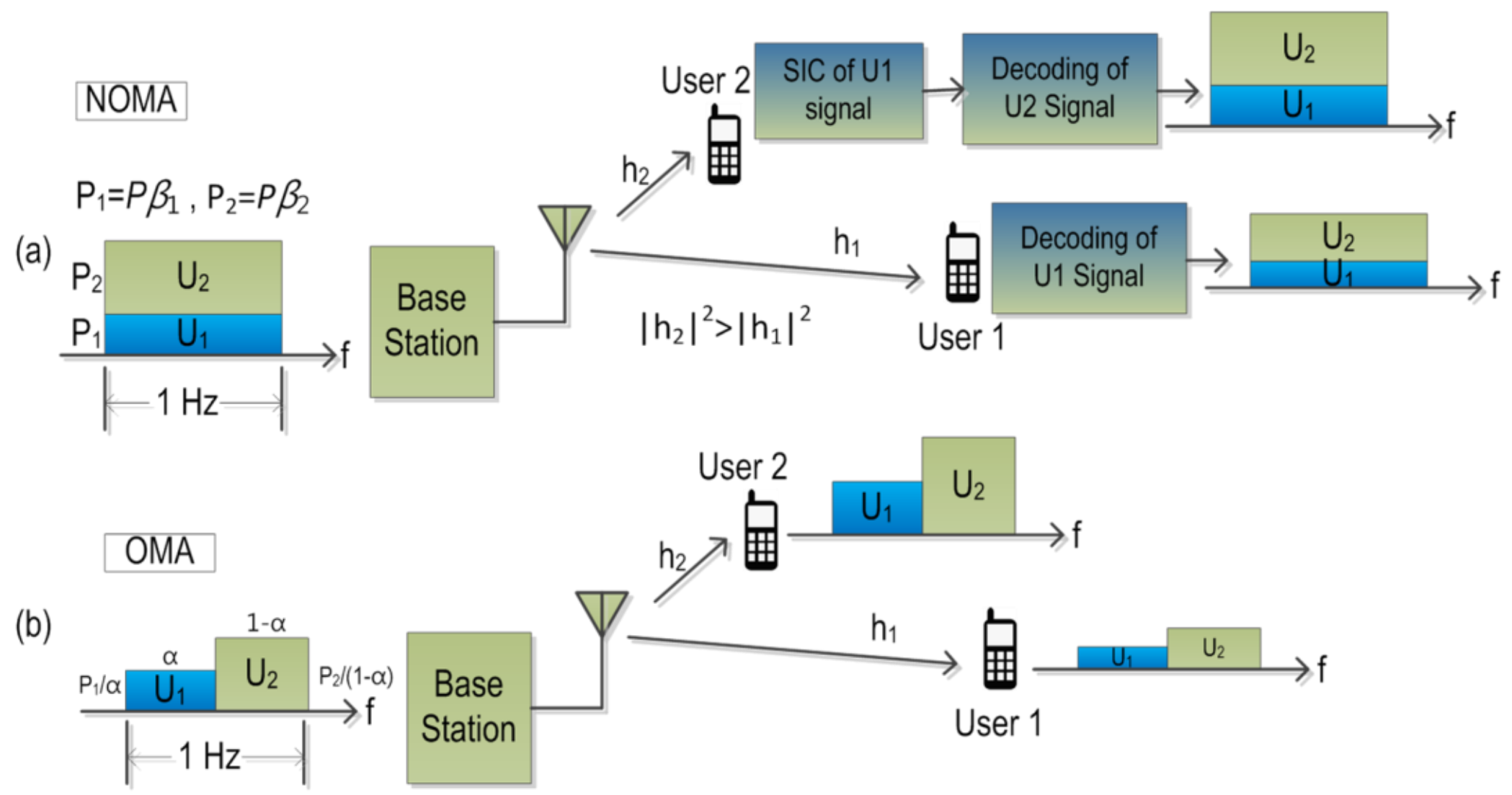

Fig. 5. Multiple access schemes for a two-user scenario (a) NOMA (b) OMA.

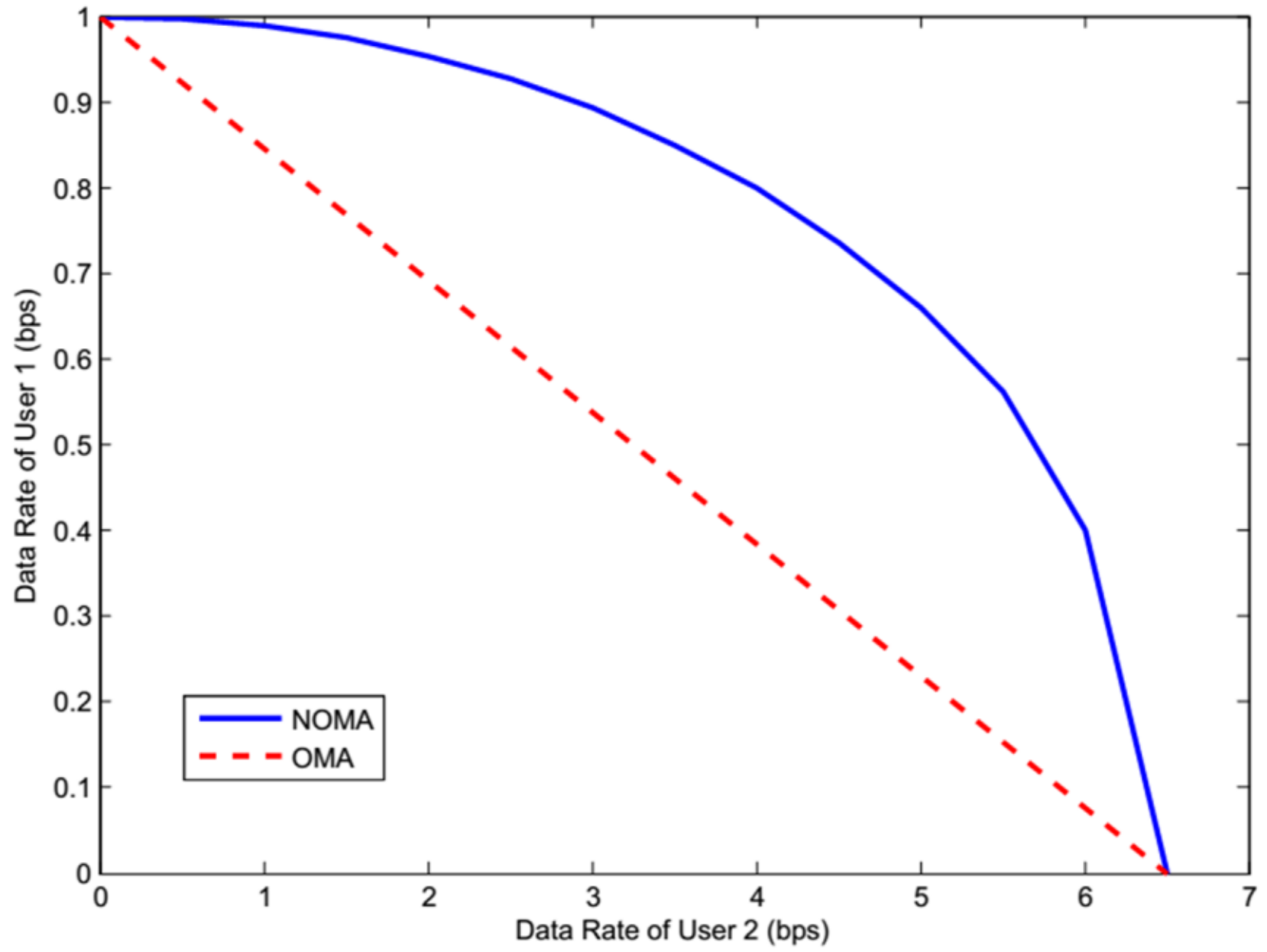

Fig. 6. Capacity regions of NOMA and OMA for downlink.

## III. Potential NOMA Solutions

This section presents an overview of the present and emerging NOMA research, in a categorized fashion, considered as potential solutions to problems or issues associated with the integration of NOMA in 5G. Detailed explanations and mathematical derivations of the techniques will be avoided, since our major focus is to cover the core ideas of the state-of-the-art NOMA research in 5G systems. Interested readers are referred to the original articles for greater depth.

### *A. Impact of Path Loss*

A substantial number of researchers have investigated the performance of NOMA schemes to study the feasibility of adopting this technique as a multiple access scheme for 5G systems. A survey by Higuchi and Benjebbour and the references therein demonstrate that NOMA can be a promising power-domain user multiplexing scheme for future radio access [19]. In a cellular network with randomly deployed users, the path loss performance of NOMA can be evaluated under two situations. In the first scenario, each user has a targeted data rate determined by the assigned quality of service (QoS). Here, the outage probability is an ideal performance metric, since it measures the capability of NOMA to meet the users' QoS requirements. In the other scenario, users' rates are opportunistically allocated according to the channel conditions. In this situation, the achievable ergodic sum rate can be investigated to evaluate NOMA performance. According to Ding et al. [20], if users' data rates and assigned power are chosen properly, NOMA can offer better outage performance than other OMA techniques. This study also showed that NOMA can achieve a superior ergodic sum rate. If the SNR is high, the outage probability of the $i$th user in a typical disk-shaped cell with radius $R_D$ can be given by:

$$P_i^{out} = \frac{\tau_i}{i}\eta^i(\psi_i^*)^i, \tag{10}$$

where $\tau_i = \frac{N!}{(i-1)!(N-i)!}$, and $\eta = \frac{1}{R_D}\sum_{l=1}^{L}\beta_l$ with $\beta_l = \frac{\pi}{l}\sqrt{1-\theta_l^2}\left(\frac{R_D}{2}\theta_l + \frac{R_D}{2}\right)\left(1+\left(\frac{R_D}{2}\theta_l + \frac{R_D}{2}\right)^{\alpha}\right)$, and $\theta_l = cos\left(\frac{2n-1}{2L}\pi\right)$. Symbols $L$, $\alpha$, and $\psi_i^*$ represent the complexity-accuracy trade-off parameter[2], the path-loss factor, and the maximum SNR corresponding to the data rate of the $i$th user, respectively. Also, with a sufficient number of users, $N$, and adequate transmit SNR, $\rho$, NOMA can achieve the following ergodic sum rate:

$$R_{erg} = \log_2(\rho \log_2\{\log_2 N\}). \tag{11}$$

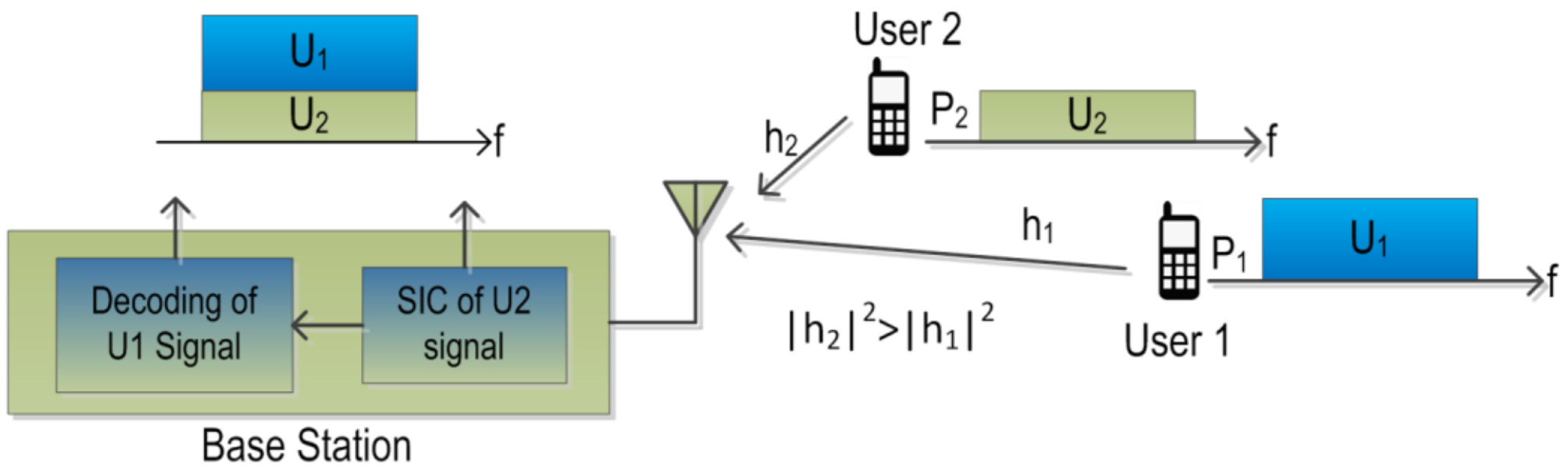

Fig. 7. NOMA in an uplink scenario.

According to the feedback framework of LTE, each user measures the downlink channel by employing reference signals and feeds back CSI in the form of predesigned transmission formats. This CSI feedback contains a rank indicator (RI), a precoding matrix indicator (PMI), and a channel quality indicator (CQI) [21]. Moreover, the rank is reported by each user at a certain time interval (i.e., semi-static). The BS can then use this feedback information for various purposes, including power control. During the channel measurement, full transmission power of BS and no interference from intra-cell users are considered by each user. On that, the reported RI is suitable for OMA. In case of NOMA, however, the transmission power is split between strong and weak users. Additionally, inter-user interference occurs in NOMA. Thus, both signal and interference powers experienced by NOMA users change, which may result in a different rank from that of reported RI assuming full transmission power. Therefore, the rank feedback in NOMA will inherently place a limitation on the achievable gain. A couple of rank optimization methods [21] could be deployed in order to overcome this limitation, thereby enhancing both the outage probability and ergodic capacity performance.

[2] Complexity-accuracy trade-off parameter: as the name suggests, the accuracy increases at the expense of increased computational complexity (in outage calculation). Therefore, a trade-off between complexity and accuracy is required in order to calculate the outage probability. The calculations of outage probabilities for NOMA users might be frequently needed for precise resource allocation.

*B. Cooperative NOMA (C-NOMA)*

In wireless networks, cooperative communications has gained a great deal of attention due to the ability to offer spatial diversity to mitigate fading, while resolving the difficulties of mounting multiple antennas on small communications terminals [22]. In cooperative communications, several relay nodes are assigned to assist a source in forwarding information to the respective destinations. Therefore, the integration of cooperative communications with NOMA can further improve system efficiency in terms of capacity and reliability. The cooperative NOMA (C-NOMA) scheme [23] exploits prior information available in NOMA systems. In this scheme, users with better channel conditions decode the messages for the others, and therefore, these users act as relays to improve reception reliability for users with poor connections to the base station. Cooperative communications for users with better channel conditions than others can be implemented by using short-range communications techniques, such as ultra-wideband (UWB) and Bluetooth (BT). In particular, C-NOMA consists of two phases, namely, transmission phase and cooperative phase. During the transmission phase, the BS sends superposed messages ($N$ users' signal) to NOMA users. At the end of this phase, successive detection will be carried out by the users. The cooperative phase consists of $(N-1)$ time slots. At the $i$th time slot, $1 \leq i \leq (N-1)$, the $(N-i+1)$th user broadcasts the combination of the $(N-1)$ messages. One can note that the power allocation coefficients at each time slot are different based on local channel conditions. This demonstrates that C-NOMA can achieve the maximum diversity gain for all users. The cooperative NOMA scheme ensures that the $i$th best user experiences a diversity of the order of $N$ conditioned on a specific power allocation ratio. However, C-NOMA is expensive in terms of additional time slots, since its cooperative phase requires message retransmissions from each user acting as a relay in a serial manner. To reduce system complexity, C-NOMA performs user pairing based on distinctive channel gains. The performance of C-NOMA can be further enhanced by adopting optimal power allocation schemes [24, 25]. Direct derivation of the theoretical achievable data rate in NOMA is quite difficult. However, if the rate of conventional TDMA is compared with that of non-cooperative NOMA, the performance difference is seen not as a function of power allocation coefficients but rather as depending on how disparate two users' channels are. And a similar observation can be noted for C-NOMA.

NOMA for a multiple-antenna relay network has been studied in [26, 27]. These studies analyzed the outage behavior of the mobile users and derived closed-form expressions for the exact outage probability. If NOMA is combined with a multiple-antenna amplify-and-forward (AF) relay network, where the base station and mobile users are equipped with multiple antennas, the relay locations have a substantial impact on the outage performance. When the relay location is close to the BS, NOMA outperforms conventional OMA. However, conventional OMA attains better outage performance when the relay location is close to the users. In either case, NOMA offers better performance in terms of SE and user fairness. Unlike C-NOMA systems, a cooperative relaying system (CRS) using NOMA [28] for spatially multiplexed transmissions enhances SE. In this system, the source transmits a superposed signal to the relay and the destination during the first time slot. During the second time slot, only the relay transmits the decoded symbol to the destination. Note that the destination receives a single data symbol during two time slots in conventional CRS (cooperative relaying without NOMA). With $P_{SR}$ and $P_{RD}$ being the average powers for the channels of source-to-relay and relay-to-destination, respectively, the average sum rate of CRS using NOMA can be obtained by:

$$R_{avg} = \frac{1}{2}\log_2\rho - \frac{\log_2 e}{2} E_c - \frac{1}{2}\log_2\left(\frac{1}{P_{SR}} + \frac{\beta_2}{P_{RD}}\right), \tag{12}$$

where $E_c$ is the Euler constant. CRS using NOMA achieves more SE than conventional CRS when the SNR is high and the average channel power of the source-to-relay link is superior to the source-to-destination and relay-to-destination links. The concept of relaying can also be adopted to extend the cell coverage or overcome shadowing in case of NOMA transmission [29] where a BS can directly communicate with a nearby user while communicating with a distant user only through a relay. In this system, the BS transmits a superposed signal to the relay and the nearby user during the first time slot. During the second time slot, while the relay transmits the decoded symbol for the distant user, the BS transmits the information for only the nearby user. Note that the nearby user can remove the interference signal that originated from the relay by using the side information (the signal of the distant user) obtained during the first time slot. The utilization of NOMA in this sort of coordinated direct and relay transmission provides considerable performance gain compared with NOMA in uncoordinated direct and relay transmission.

### *C. Fairness in NOMA*

NOMA users experience unequal data rates, since this access method is based on SIC decoding order [2, 3]. Although the decoding order is formulated based on users' CSI, the NOMA service could be critical for some situations where strict fairness might be an issue. The power allocation problem from a fairness viewpoint can be investigated under two assumptions: i) the BS has perfect CSI, and hence, users' data rates adapt to the channel conditions; and ii) users have fixed targeted data rates under an average CSI. With these assumptions, it is possible to come up with low-complexity algorithms that yield globally optimal solutions. NOMA with a fairness constraint also outperforms OMA approaches by significantly improving the performance of the users with the worst channel conditions [30]. If instantaneous CSI is available at the BS, fairness among users can be ensured by maximizing the minimum achievable data rate, i.e.,

$$\max_{\beta} \min_{i \epsilon N} R_i(\beta), \tag{13}$$

$$\text{s.t.} \sum_{j=1}^{N} \beta_j = 1, \tag{13a}$$

$$0 \leq \beta_j, \quad for\ j \epsilon N. \tag{13b}$$

Since problem (13) is not convex, it needs to first be converted into a sequence of linear programming. Eventually, the optimal solution to (13) can be given by:

$$\beta_i = \frac{2^t - 1}{P|h_i|^2}\left(P|h_i|^2 \sum_{k=i+1}^{N} \beta_k + \sigma_n^2\right), \quad i = N, N-1, \ldots, 1, \tag{14}$$

where $t$ represents the minimum data rate. If instantaneous CSI is not available, the outage probability should be optimized with average CSI. In this case, fairness among users can be ensured by minimizing the maximum outage probability as $\min_\beta \max_{i\epsilon N} P_i^{out}(\beta)$ conditioned on (13a) and (13b). The fairness of a NOMA system can also be determined by controlling the rate allocated to concurrent transmissions, which in turn can be achieved by using different scheduling approaches [31].

### *D. NOMA with MIMO and Beamforming*

Beamforming (BF) is a signal processing technique used in various wireless systems for directional communications. Multi-user BF in MIMO systems is known as a capacity-enhancing technology. In a multi-user BF system, each user is supported by a single BF vector, orthogonal to the other users' channels in order to eliminate interference from other users and thus maximize the achievable sum capacity. The integration of NOMA with multi-user BF (NOMA-BF) thus has the potential to capture the benefits of both NOMA and BF [25]. The NOMA-BF technique allows two users to share a single beamforming vector. To reduce the inter-beam interference (from users of other beams) and intra-beam interference (from users sharing the same beamforming vector), NOMA-BF comes with a clustering and power allocation algorithm based on correlation among users and channel gain difference, respectively. The NOMA-BF system improves the sum capacity, compared to the conventional multi-user beamforming system. NOMA-BF also guarantees weak users' capacity to ensure user fairness. Suppose there is a two-user NOMA-BF consists of $N$ clusters, where each of them contains two users. In this scenario, a power allocation scheme for the $n$th cluster maximizes the sum capacity while keeping the weak user's capacity at least equal to that of the conventional multi-user beamforming system. It can be formulated as follows, conditioned on (13a) and (13b):

$$\beta_1^n = \arg\max_{\beta_1^n}(R_1 + R_2), \tag{15}$$

$$\text{s.t.}\, R_2 \geq \frac{1}{2} R_{2,conv-BF}, \tag{15a}$$

where $R_1$ and $R_2$ are the capacities of the strong and weak users, respectively, whereas $R_{2,conv-BF}$ is the capacity of the weak user if the weak user is supported by conventional beamforming. Symbols $\beta_1^n$ and $\beta_2^n = 1 - \beta_1^n$ are the power fractions of the strong and weak users, respectively, in the $n$th cluster. The optimal solution to (15) can be obtained by using the Karush-Kuhn-Tucker (KKT) condition below:

$$\beta_1^n = \frac{1}{\sqrt{\left(1+\left|h_{2,n}\right|^2\rho\right)}} - \frac{\left\{\sqrt{\left(1+\left|h_{2,n}\right|^2\rho\right)}-1\right\}\left\{\sum_{i=1,i\neq n}^{N}\left|h_{2,i}w_i\right|^2\rho+1\right\}}{\rho\left|h_{2,i}w_i\right|^2\sqrt{\left(1+\left|h_{2,n}\right|^2\rho\right)}}. \quad (16)$$

The above clustering algorithm, based on channel correlation and a zero forcing (ZF) precoding matrix, helps two users in the same cluster to achieve similar benefits from the ZF precoding matrix. However, the performance of the system becomes considerably degraded due to the increased multi-user interference (MUI) when there is a shortage of user sets with high channel correlation. In this regard, researchers investigated the design of the precoding matrix for MU-MIMO NOMA to achieve a higher sum-rate in the presence of said interference [32].

As discussed above, random opportunistic beamforming [25] for the MIMO NOMA system works under the assumption of perfect CSI at the transmitter. It becomes evident that with a relatively large number of users, the combination of NOMA and MIMO can achieve sufficient throughput gain [33]. If perfect CSI at the transmitter is unavailable due to limited feedback, statistical CSI can be utilized for long-term power allocation to maximize the ergodic capacity of MIMO NOMA systems. The systematic implementation of MIMO NOMA has been reported in [21, 24, 34] and demonstrates that the use of MIMO NOMA can outperform conventional MIMO OMA. Both optimal and low-complexity suboptimal power allocation schemes [35] can maximize ergodic capacity with a total transmit power constraint. Ding et al. proposed a new design for precoding and detection matrices in a general NOMA downlink scenario with a fixed power allocation scheme for all the participating users [36]. The analytical and numerical results confirmed that MIMO NOMA can achieve better outage performance than conventional MIMO OMA, even for users who experience strong co-channel interference. Suppose a BS equipped with $M$ antennas communicates with several users equipped with $N$ antennas each. Also, the users are randomly grouped into $M$ clusters with $K$ users in each cluster. Then, the outage probability, at a high SNR, for the $k$th ordered user in the $i$th cluster can be given by

$$P^{out} \approx \frac{K!\left[\frac{(\epsilon^*)^{N-M+1}}{(N-M+1)!}\right]^{K-k+1}}{(K-k)!\,(k-1)(K-k+1)}, \quad (17)$$

where $\epsilon^*$ represents the maximum required SINR to meet the QoS of users in the $i$th cluster. As can be seen from (17), MIMO NOMA can experience a diversity gain of $(N - M + 1)(K - k + 1)$. It is also intuitive that the extension of NOMA in massive MIMO systems can further enhance the SE. Unlike the single-antenna system, where the strong user is responsible for successful SIC decoding, precoders in MIMO systems usually affect the signal-to-interference-plus-noise ratio (SINR) of both users. Thus, the weak user data rate will be limited by not only the decoding performance of the weak user but also by the decoding capability of the strong user. In this respect, the same precoder for both users in each cluster may not be sufficient. The above work on MIMO NOMA deals with the same precoder in each cluster.

However, it is possible to further enhance the sum rate if the two users in a cluster can use different precoders [37].

*E. NOMA in a Coordinated System*

In cellular systems, a cell-edge user usually experiences a lower data rate compared to that experienced by a user near the BS. Coordinated multipoint (CoMP) transmission (and reception) techniques, where multiple BSs support cell-edge users together, are usually employed to increase transmission rates to cell-edge users. The associated BSs for CoMP need to allocate the same channel to a cell-edge user. As a result, the SE of the system worsens as the number of cell-edge users increases. In order to avoid this problem, a coordinated SC (CSC)-based NOMA scheme [38] was formulated by considering SC for downlink transmissions to a group of cell-edge users and users near the BS simultaneously with a common access channel [17]. In other words, BSs transmit Alamouti (space-time) [39] coded signals to user $c$ (a cell-edge user), while each BS also transmits signals to a user near the BS. The CSC-NOMA scheme with Alamouti code provides a cell-edge user with a reasonable transmission rate without diminishing the rates to nearby users, and increases SE. If $R_{c1}$, $R_{c2}$, and $R_c$ are the rates to a user ($U_1$) near BS 1, a user ($U_2$) near BS 2, and a coordinated user ($U_c$), respectively, the sum rate becomes $R_{c1} + R_{c2} + R_c$ with

$$R_{c1} = E\left[\log_2\left(1 + \frac{|h_{1,1}|^2 P_1}{E\left[|h_{1,2}|^2\right] P_2 + \sigma_n^2}\right)\right], \tag{18}$$

$$R_{c2} = E\left[\log_2\left(1 + \frac{|h_{2,2}|^2 P_2}{E\left[|h_{2,1}|^2\right] P_1 + \sigma_n^2}\right)\right], \tag{19}$$

$$R_c = \min\{Z_1, Z_2, Z_c\}, \tag{20}$$

where $Z_1 = E[\log_2(1 + SINR_1)]$, $Z_2 = E[\log_2(1 + SINR_2)]$, and $Z_c = E[\log_2(1 + SINR_c)]$. Let $SINR_i$ be the SINR for user $U_i$ in decoding the signal of user $U_c$. Note that $h_{i,j}$ denotes the channel coefficient from the BS $j$ to user $U_i$. Because of the use of Alamouti code for CoMP communications, the exchange of instantaneous CSI is not required. This is a significant advantage over coherent transmission schemes that require instantaneous CSI exchange, which results in an excessive backhaul overhead for high mobility cell-edge users.

*F. Network NOMA*

Suppose there is a simple two-cell scenario of a cellular system (see Fig. 8), where $U_3$ and $U_4$ are served by BS 1, while $U_1$ and $U_2$ are served by BS 2. Also, assume that a two-user NOMA scheme is adopted, so that $U_3$ is paired with $U_4$, and $U_1$ is paired with $U_2$. As a cell-edge user in NOMA does not perform SIC before decoding its signal, $U_4$ and $U_1$ are inherently unable to avoid the interference from $U_3$ and $U_2$, respectively. In order to reduce the impact of this interference,

BS 1 and BS 2 allocate more power to $U_4$ and $U_1$, respectively. As such, inter-cell interference occurs in downlink transmissions and mutual interference between $U_4$ and $U_1$ occurs in uplink transmissions [40]. To deal with this problem, associated with the employment of NOMA in a multi-cell scenario, the straightforward application of single-cell NOMA solutions will not be appropriate; single-cell NOMA needs to be extended to network NOMA. Like various inter-cell interference mitigation techniques employed in OMA, some solutions are also required in NOMA to reduce the impact of interferences in the context of a multi-cell scenario. One possible solution to mitigate the aforementioned interferences in network NOMA is to utilize joint precoding of users' signals across the neighboring cells. However, the design of an optimal precoder is difficult, since each BS should know all users' data and CSI. The correlation-based precoder design needs dynamic user selection for each NOMA pair [25]. Moreover, the multi-user precoding applicable to single-cell NOMA may not be realistic in network NOMA, since a beam generated via geographically separated BSs does not support more than one spatially separated user for intra-beam NOMA. A low-complexity precoding scheme for network NOMA was introduced [40] based on the fact that large-scale fading would be very disparate between the links to different cells. Here, the joint precoder is applied only to cell-edge users (e.g., $U_4$ and $U_1$ in Fig. 8) and the resulting SINR of each user $U_i$ of power $P_i$ is found as:

$$SINR_1 = \frac{\left[\{(\boldsymbol{H}_{41}(\boldsymbol{H}_{41})^{\boldsymbol{H}})^{-1}\}_{1,1}\right]^{-1} P_1}{|h_{11}|^2 P_0 + |h_{12}|^2 P_2 + N_0 B}, \tag{21}$$

$$SINR_2 = \frac{|h_{22}|^2 P_2}{|h_{21}|^2 \left(P_0 + \left|w_{0,0}\right|^2 P_1 + \left|w_{0,1}\right|^2 P_3\right) + N_0 B}, \tag{22}$$

$$SINR_3 = \frac{|h_{30}|^2 P_3}{|h_{32}|^2 \left(P_2 + \left|w_{1,0}\right|^2 P_1 + \left|w_{1,1}\right|^2 P_3\right) + N_0 B}, \tag{23}$$

$$SINR_4 = \frac{\left[\{(\boldsymbol{H}_{41}(\boldsymbol{H}_{41})^{\boldsymbol{H}})^{-1}\}_{0,0}\right]^{-1} P_4}{|h_{41}|^2 P_0 + |h_{42}|^2 P_2 + N_0 B}, \tag{24}$$

where $\boldsymbol{H}_{41} = [\boldsymbol{h}_4, \boldsymbol{h}_1]^T$ with the channel vector of the $i$th user, $\boldsymbol{h}_i = [h_{i1}, h_{i2}]$, and $h_{ij}$ ($i \in \{1, 2, 3, 4\}$, and $j \in \{1, 2\}$) being the channel response between the $j$th BS and $i$th user. The zero-forcing precoder, $W$, is the normalized pseudo-inverse of $\boldsymbol{H}_{41}$, i.e., $W = (\boldsymbol{H}_{41})^H(\boldsymbol{H}_{41}(\boldsymbol{H}_{41})^{\boldsymbol{H}})^{-1}$, $B$ is the system bandwidth, and $N_0$ is noise spectral density.

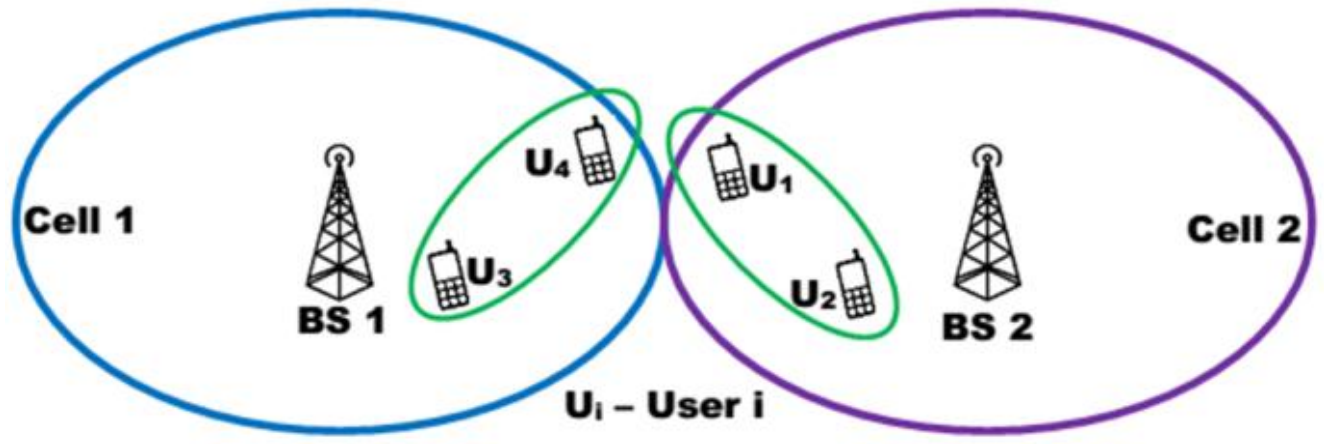

Fig. 8. NOMA in a two-cell scenario (network NOMA).

### *G. NOMA User Pairing*

Since NOMA is an interference-limited system, it is practically unwise to ask all users in the system to perform NOMA jointly. In this regard, users can be divided into multiple groups, where NOMA is implemented within each group, and different groups are allocated with OMA. Clearly, the performance of this hybrid MA scheme depends on which users are grouped together. The impact of user pairing/grouping was investigated mathematically in [41]. This work demonstrates that the performance gain of NOMA with fixed power allocation over conventional OMA can be further increased by selecting users whose channel conditions are more distinctive. A two-step method for user pairing based on proportional fairness (PF) was introduced [42]. In the first step, the power allocation for each candidate user set is optimized to find the highest scheduling metric. In a subsequent step, the optimal user pair, or single user with the maximum metric, is scheduled. Since this user-pair power allocation (UPPA) technique avoids unnecessary comparison of candidate user pairs by formulating the prerequisites for user pairing, this method comes with considerably lower computation complexity compared to tree-search–based transmission power allocation (TTPA). Suppose, the upper and lower bounds of the allocated power coefficient of $U_1$ are $\beta_1^u$ and $\beta_1^l$, respectively. Now, if any of the following conditions, termed the prerequisites, are not satisfied by a candidate user pair, the computation and comparison of their PF scheduling factor can be omitted. So the compared number of user pairs can be substantially reduced.

$$\beta_1^u > 0 \quad \Leftrightarrow \quad I_2 > \rho_0 > 0, \tag{25}$$

$$\beta_1 \geq \rho_0 I_1^{-1} \triangleq \beta_1^l, \tag{26}$$

$$\beta_1^l < 1 \quad \Leftrightarrow \quad I_1 > \rho_0 > 0, \tag{27}$$

where $I_i$ is the channel quality indicator (CQI) of user $U_i$, and $\rho_0$ is the minimum required SINR threshold, while the modulation and coding scheme (MCS) with the lowest coding rate is selected for an expected block error rate. Eventually, the solution to the optimal power ratio provided by the two-user proportional fairness-based NOMA can be given by:

$$\hat{\beta}_1 = \begin{cases} \beta_1^* & \beta_1^l \leq \beta_1^* \leq \beta_1^u \\ \beta_1^u & R_1 < R_2 \\ \beta_1^l & others \end{cases}, \tag{28}$$

where $\beta_1^*$ is the solution to the modified scheduling factor [43]. As (28) indicates, the available scheduling factor of every valid pair of users can be computed, and the largest one can be selected. A similar user set–selection algorithm is formulated based on the mathematical characteristics of the PF metric [44]. This method can also reduce the computational complexity by judging whether a user set is worth multiplexing based on a simple condition between the weakest two users within the set. Also, the impact of user pairing on the performance gain of NOMA with a MIMO system was investigated [36], demonstrating that the performance gain of MIMO NOMA over MIMO OMA increases as the number of users in each group increases.

*H. Energy-Efficient NOMA*

NOMA employs some controllable interference via non-orthogonal resource allocation and realizes overloading at the cost of slightly increased receiver complexity. Consequently, higher SE can be achieved by NOMA for 5G. Although SE shows how efficiently a limited spectrum resource is utilized, it fails to provide any insight on how efficiently energy is utilized. With the rise in desire for green communications in recent years, reducing energy consumption has become of prime importance for researchers, and 5G has also targeted EE as one of the major parameters to be achieved. Nonetheless, Shannon's information capacity theorem illustrates that the two objectives of minimizing consumed energy and maximizing SE are not achievable simultaneously, and calls for a trade-off. It can be noted that with circuit power under consideration, there always exists an optimal point in the EE versus SE (EE-SE) curve. An energy-efficient two-user single-cell NOMA was studied in [40]. Under fixed total power consumption, the EE-SE relationship was found to be linear with a positive slope. Appropriate power allocation between two users allows achieving any point in the EE-SE curve. For a given SE for each user, maximum EE performance can be achieved. The degree of efficiency can be adjusted by varying total power using power-control schemes. If the sum rate capacity of the cell is $R_{sum}$ with total power consumption $P_{cell}$, EE can be written as $\eta_E = R_{sum}/P_{cell} = B\eta_S/P_{cell}$, where $\eta_S$ is the spectrum efficiency. The EE optimization can be achieved in both single input single output and MIMO systems [45].

*I. NOMA in Visible Light Communications*

One of the major downsides to visible light communications (VLC) systems is the narrow modulation BW of the light sources, which results in a barrier to attaining competent data rates. Like wireless communications, optical wireless communications also considers various signal-processing techniques, and multicarrier and multi-antenna systems for achieving higher data rates in VLC systems. Since NOMA is now a potential candidate for next-generation wireless communications, the feasibility of NOMA in VLC is also be a subject of interest. It is viable to apply a NOMA scheme to enhance the achievable throughput in high-rate VLC. In fact, studies reveal that NOMA is a promising MA scheme for downlink in VLC networks [46, 47]. However, these works simply reflect a scenario with multiple light emitting diodes (LEDs) in an indoor environment; real high-speed OFDM-based communications has not been considered. Other research investigated optical OFDM transmission and compared the achievable capacities of NOMA and OFDMA [48]. It also analyzed the impact of cancellation error in SIC receivers for VLC systems. The achievable data rates for $U_1$ and $U_2$ in two-user NOMA with DC-biased optical OFDM (DCO-OFDM), assuming a frequency-flat fading channel, can be written as

$$R_1 = \frac{B}{2}\log_2\left(1 + \frac{\gamma_o^2 L_o^2 P \beta_{1,n} G_1^2}{\epsilon_o \gamma_o^2 L_o^2 P^2 (1 - \beta_{1,n}) G_1^2 + N_0 B}\right), \tag{29}$$

$$R_2 = \frac{B}{2}\log_2\left(1 + \frac{\gamma_o^2 L_o^2 P \beta_{2,n} G_2^2}{\gamma_o^2 L_o^2 P^2 \beta_{2,n} G_2^2 + N_0 B}\right), \tag{30}$$

where the optical parameters $(\gamma_o, L_o, \epsilon_o)$ are receiver responsivity, number of LED chips, and the signal cancellation error, symbols $G_1$ and $G_2$ are channel gain for $U_1$ and $U_2$, respectively, and the symbol $\beta_{i,n}$ denotes the power coefficient for the $n$th subcarrier for $U_i$.

### *J. NOMA with Link Adaptation*

Hybrid automatic repeat request (HARQ) protocol, an indispensable part of link adaptation, is designed for reliable communications by using retransmission diversity and channel coding gain. Suppose, when multiple NOMA packets are transmitted, there is a collision, and a retransmission request is sent to the users in collision. Each packet can convey at least a certain amount of information, even though the packets collided. All the retransmitted signals can be combined by using HARQ in order to improve the SE of NOMA [49]. Li et al. provided a thorough investigation of HARQ design for NOMA with single user-MIMO (SU-MIMO) [50]. They found that the HARQ technique for a NOMA system is more challenging than for an OFDMA system. A good HARQ algorithm should deal with two central problems in NOMA retransmission: when the HARQ combination should be directed, and how to conduct power assignment for retransmission. Based on channel responses to initial transmission and retransmission, it was found that the optimum time for conducting an opportunistic HARQ combination by the distant user is when the SINR of the initial transmission is reasonably larger than the retransmission SINR. And one approach for conducting retransmission power assignment would be assigning the power in such way so that it maximizes the geometric mean user throughput of all the users in retransmission. In sum, an advanced HARQ design can bring NOMA gain improvement both in cell throughput and cell-edge throughput.

### *K. Other NOMA Solutions*

- NOMA with Raptor Codes

For a given integer, $k$, and a real number $\epsilon$, Raptor codes, first proposed by Shokrollahi [51], encode a message of $k$ symbols into a potentially limitless sequence of encoding symbols such that any subset of $k\ (1+\epsilon)$ encoding symbols allows the message to be recovered with high probability. Raptor codes were recently found effective in several cooperative communications scenarios. The integration of Raptor codes with NOMA has been studied in [52], where an interfering channel with Raptor code was added to an existing main non-orthogonal wireless channel. It was demonstrated that the coded interference does not affect the performance of the main channel, while the interfering signal itself can successfully be decoded with high probability.

- NOMA with Network Coding

Random linear network coding (RLNC) is a good encoding scheme that allows data retransmission. In RLNC, the source does not need to be aware of the packets lost by the intended receiver. To date, various RLNC techniques have been proposed to improve transmission efficiency in both multicast and broadcast services. The performance of multicast services in downlink networks can be further enhanced by integrating RLNC with NOMA. NOMA with RLNC was studied by Hagh and Soleymani in [53]. In conventional NOMA, the power domain multiplexing of multiple receivers is considered for unicast services, whereas NOMA-RLNC utilizes power-domain multiplexing for multiple reception groups of receivers for multicast services. It was found that NOMA-RLNC improves the packet success probability of providing multicast services where a source superposes multiple coded packets before transmitting.

- Coexistence of NOMA and OMA

In terms of capacity enhancement, which is a major goal of 5G, NOMA is a potential candidate for future radio access. That said, it does not mean OMA schemes will be entirely replaced by NOMA. For example, OMA might be preferred over NOMA for small cells if the number of users is small and the near–far effect is not important. It can be concluded that both OMA and NOMA will coexist to fulfill varied requirements of different services and applications in future 5G. As a matter of fact, the long-term coexistence of different radio access technologies is, in general, an important feature of 5G networks. Dai et al. discussed NOMA schemes for 5G and analyzed their basic principles, key features, and receiver complexity [15]. They conclude that the concept of software-defined multiple access can offer various services and applications with different requirements.

- SIC Receiver Variations

An ideal SIC receiver perfectly cancels the interference from the cell-edge user at the receiver of a cell-center user. Until now, the seminal works on NOMA adopt a SIC receiver at cell-center users, while cell-edge users do not take SIC detection into account. However, it is possible to further improve the performance of NOMA by considering an advanced receiver at cell-edge users [54]. By advanced receiver, this paper means a receiver that can perform ideal SIC processing, since the improvement in NOMA performance gain with non-ideal SIC is not satisfactory. SIC where the cell-edge user’s signal is demodulated and a hard decision is made without channel decoding is called symbol-level SIC (SLSIC). In codeword-level SIC (CWSIC), the signal of a cell-edge user is demodulated and decoded. The probability of successful cell-edge user signal recovery increases, compared with SLSIC, since channel decoding is involved in signal detection. Consequently, the impact of error propagation can be reduced in CWSIC [55, 56]. However, it increases computational complexity. If the Gray mapping of superposed signals be achieved by jointly modulating multi-users’ signals at the transmitter, the desired signal can be decoded with the application of a log-likelihood method at the receiver. Since the decoding can be performed without SIC processing, the complexity of CWSIC can be reduced [55]. Appropriate power control mechanism for NOMA employing minimum mean squared error-based linear filtering followed by SIC (MMSE-SIC) can mitigate inter-cell interference and the fairness can be ensured by employing a PF-based multiuser scheduling [57].

- NOMA with SWIPT

In addition to enhancing SE, another major goal of 5G networks is to maximize EE. Many researchers demonstrated that simultaneous wireless information and power transfer (SWIPT) is expected to provide a viable solution to overcome EE issues. It is possible to establish a cooperative communications protocol by combining SWIPT and NOMA, where cell-center users act as energy-harvesting relays to help cell-edge users. The use of SWIPT does not affect the diversity orders of both nearby and distant users, compared to conventional NOMA [58].

## IV. PERFORMANCE EVALUATIONS

To illustrate the performance gains, this section provides numerical results for various NOMA setups. Two primary performance metrics, namely, outage probability and achievable rate are considered for this purpose. Also, the section provides some insights into SE and EE performances of NOMA.

### *A. Outage Probability*

Based on (10), Fig. 9 compares the outage performance of a NOMA scheme with that of an OMA scheme for a cellular network comprising randomly deployed users with $N = 2, L = 10, \alpha = 2,$ and $R_D = 3$ m. The users are uniformly located. The targeted data rates of 0.1 bits per channel use (BPCU) and 0.5 BPCU, respectively, for a weak user and a strong user (user 2) were used. Since the conventional orthogonal scheme is considered for benchmarking, its targeted rate is 0.6 BPCU (the addition of two users' data rates). Also, note that the numerical results are based on the normalized SNR model. As can be seen from this figure, NOMA outperforms the comparable scheme, and the diversity order of the users is a function of their channel conditions. Note that in this case, the ratio of the power assigned to a strong user to the power assigned to a weak user is 1:4. The outage probability given by (7) is basically valid at a high SNR condition. On that, in order to recognize the comparative outage performance, attention should be paid to high SNR regions where both users outperform the OMA scheme. Since the assigned power to the strong user is proportionally lower, the outage performance at a low SNR region is poor. However, as the SNR becomes high enough, the power-domain multiplexing becomes dominant, and thereby shows the best performance with superior diversity order.

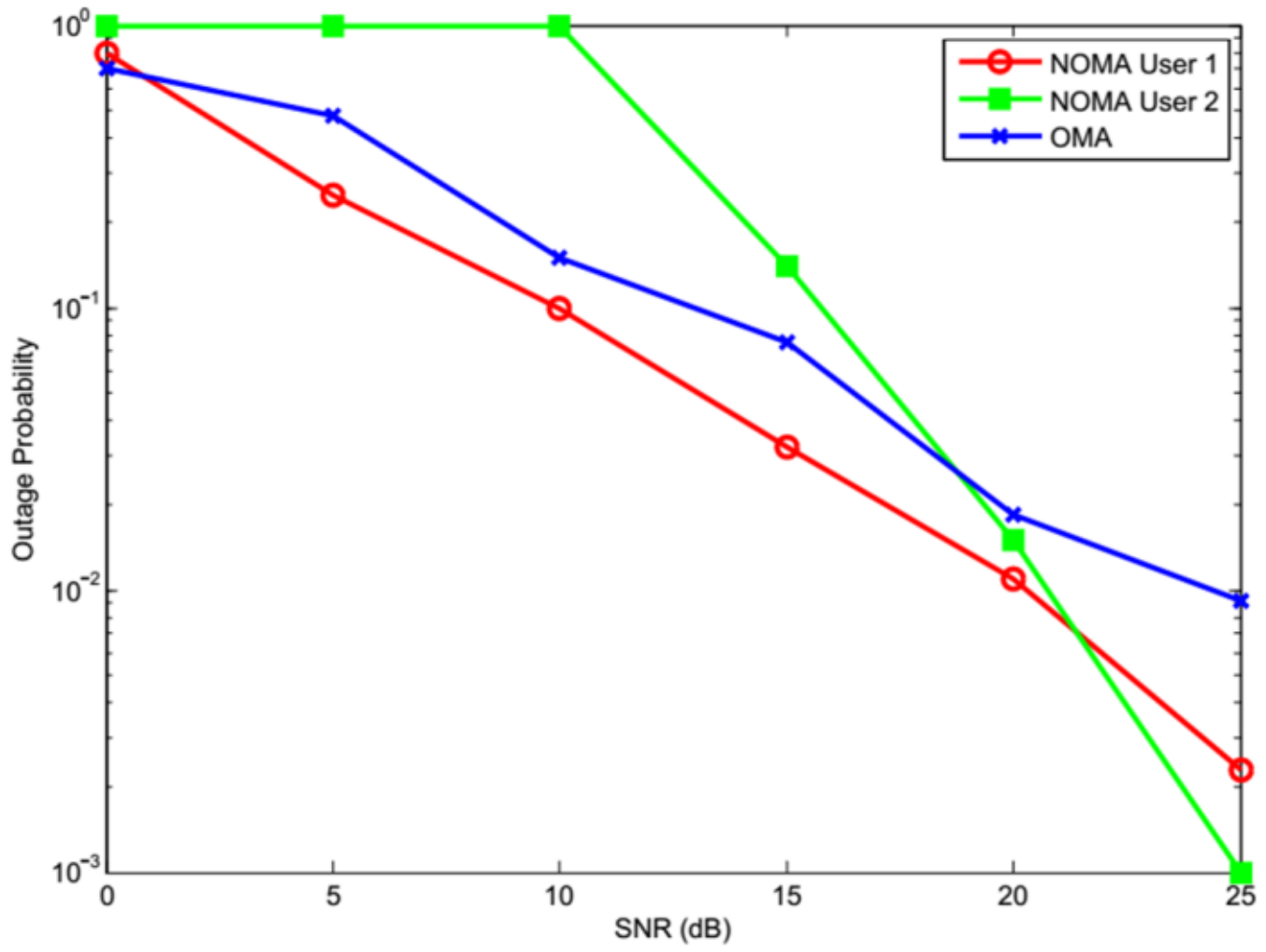

Fig. 9. Outage performance of NOMA in 5G with random users in a cell.

With the same number of users and a power allocation ratio used in Fig. 9, Fig. 10 presents the outage probability that is achieved by non-cooperative NOMA and cooperative NOMA (discussed in Section III. B) as a function of SNR. It shows that cooperative NOMA transcends the comparable scheme, since it ensures that the maximum diversity gain is achievable by all users. This high diversity gain is explained below. Under C-NOMA, users with the worst channel condition get assistance from the other $N-1$ users, along with their own direct links to the source. Although non-cooperative NOMA can attain only a diversity order of $i$ for the $i$th ordered user, C-NOMA ensures that a diversity order of $N$ is achievable for all users by exploiting user cooperation.

Fig. 11 demonstrates the performance of a typical MIMO NOMA system (discussed in Section III. D) in terms of outage probability. This figure considers two clusters, where each cluster contains two users equipped with three antennas. The power allocation ratio is 1:4, and target data rates of the strong user and the weak user are 3 BPCU and 1.3 BPCU, respectively. The figure indicates that the MIMO NOMA system outperforms MIMO with OMA, particularly at high SNR values. By comparing the slopes of the performance curves, it can be concluded that the diversity gain of MIMO NOMA is the same as MIMO OMA. However, MIMO NOMA provides better SE,

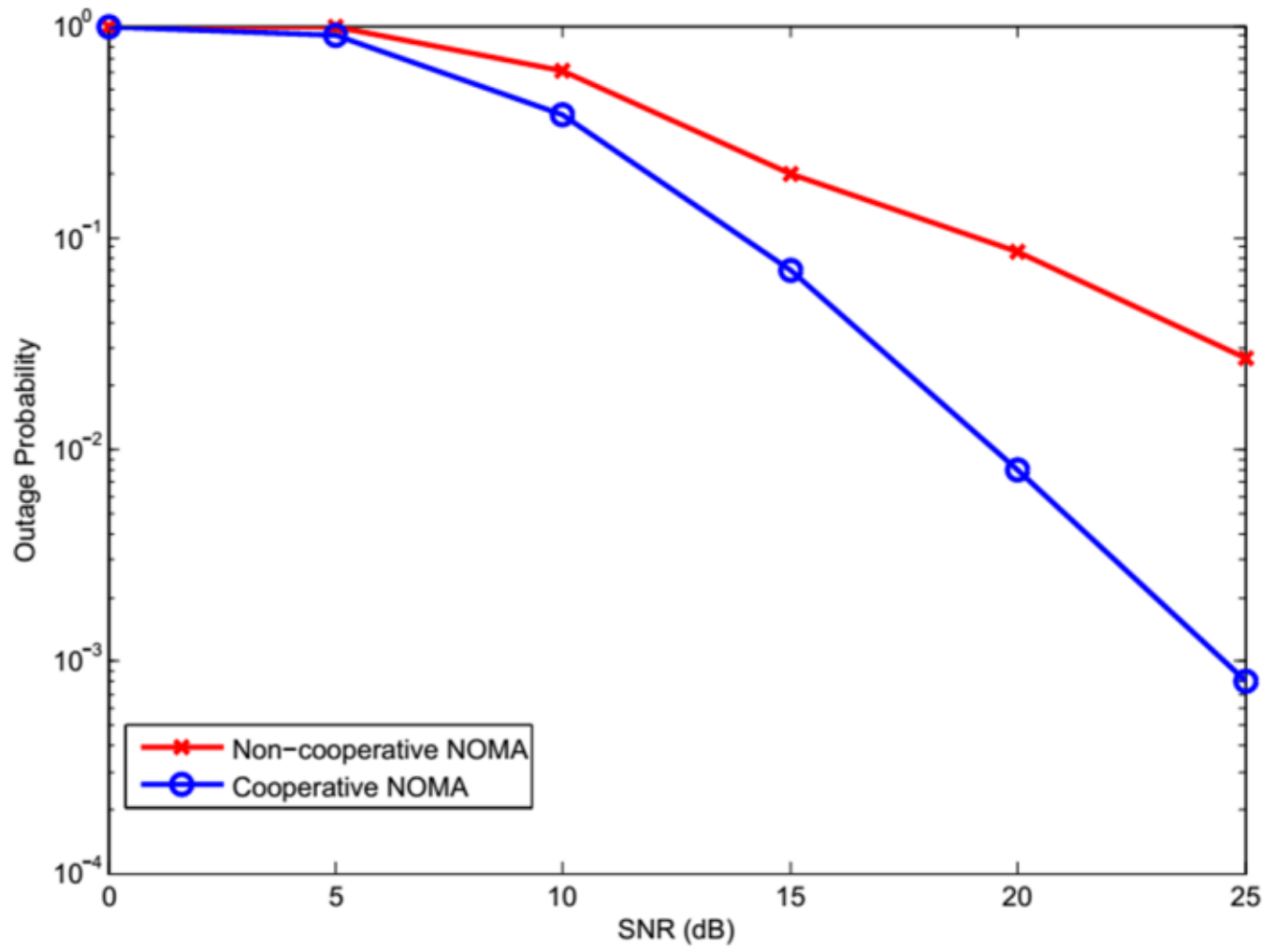

Fig. 10. Outage performance of a cooperative NOMA.

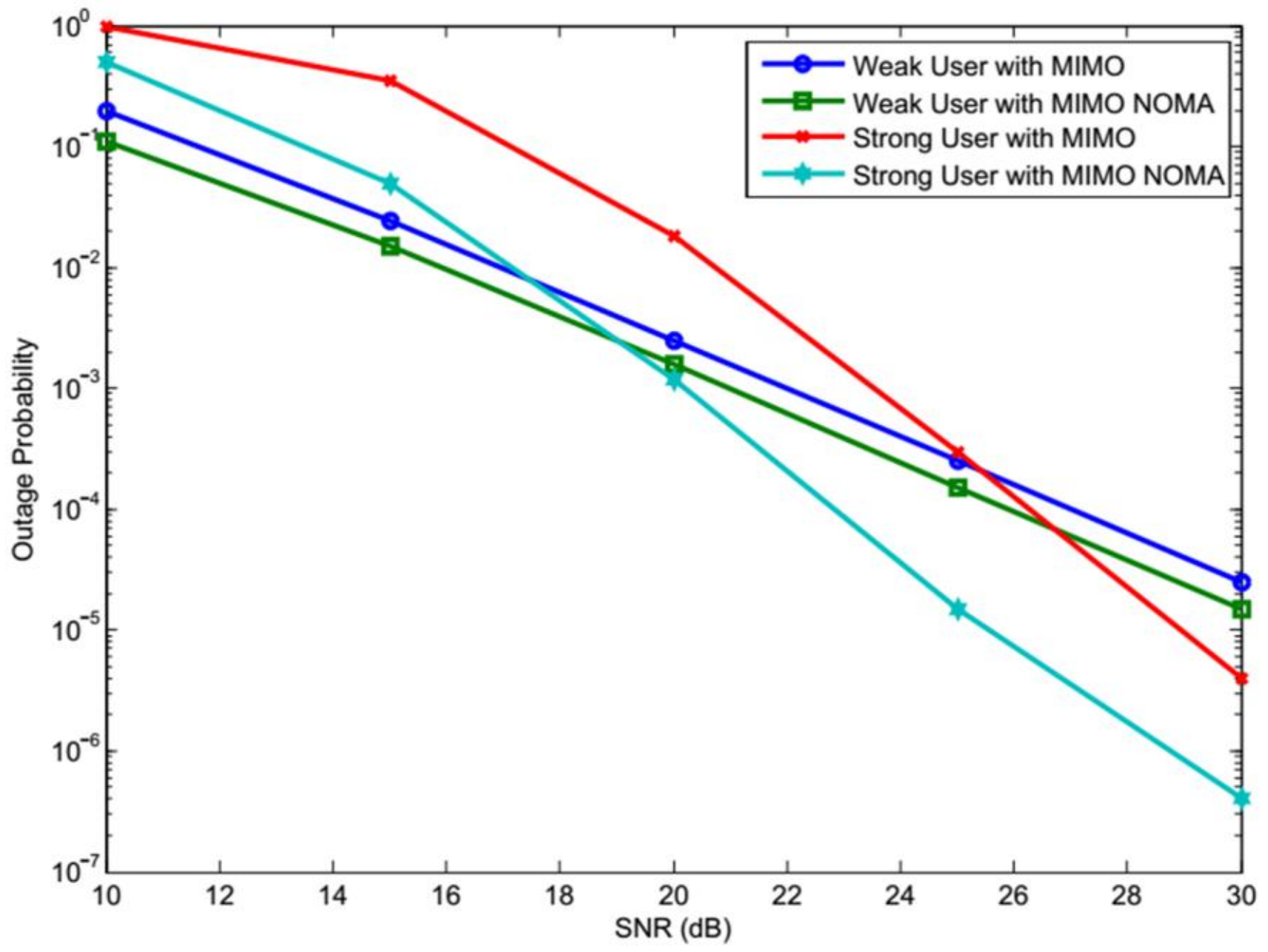

Fig. 11. Outage performance of a MIMO NOMA system.

since this gain is achieved by allowing both users from the same cluster to utilize the same BW.

### *B. Achievable Rate*

With the same system parameters used in Fig. 9, Fig. 12 presents the sum rate (the sum of the data rates of the NOMA users) achieved by a two user NOMA as a function of SNR. The targeted data rates of the randomly deployed users in the cell are determined opportunistically according to the channel conditions. In this case, the data rate of user 2 has always been considered higher than that of user 1, since the channel gain of user 2 is always higher than that of user 1. With this consideration, the figure demonstrates that NOMA can attain a higher sum rate than the OMA scheme across the entire SNR range.

In Fig. 13, the rate performance of a cooperative relaying system using NOMA (discussed in Section III. B) with fixed power allocation is presented. The figure demonstrates that CRS with NOMA achieves better average rate performance than the usual relaying system, particularly at a high SNR, when the average power of the channel coefficient of source to relay is much higher than that of relay to destination. Therefore, the relays should be nominated from the cell-center users in a NOMA-based relay network in order to get the benefits of NOMA.

Fig. 14 shows the sum capacities of NOMA-BF (discussed in Section III. D) and conventional multi-user beamforming with a correlation threshold $\rho = 0.75$, system BW of 4.32 MHz, maximum transmission power per cluster at 43 dBm, and noise density of -169 dBm/Hz. As can be seen, NOMA-BF improves the sum capacity. Here, the users are randomly located with uniform distribution in a cell radius of 500 m. NOMA-BF is better in terms of sum capacity compared to conventional multi-user beamforming, since correlation-based clustering with effective power allocation reduces the inter-beam and intra-beam interference. In view of the fact that two users share a single beamforming vector, the number of supportable users can easily be increased by utilizing NOMA-BF.

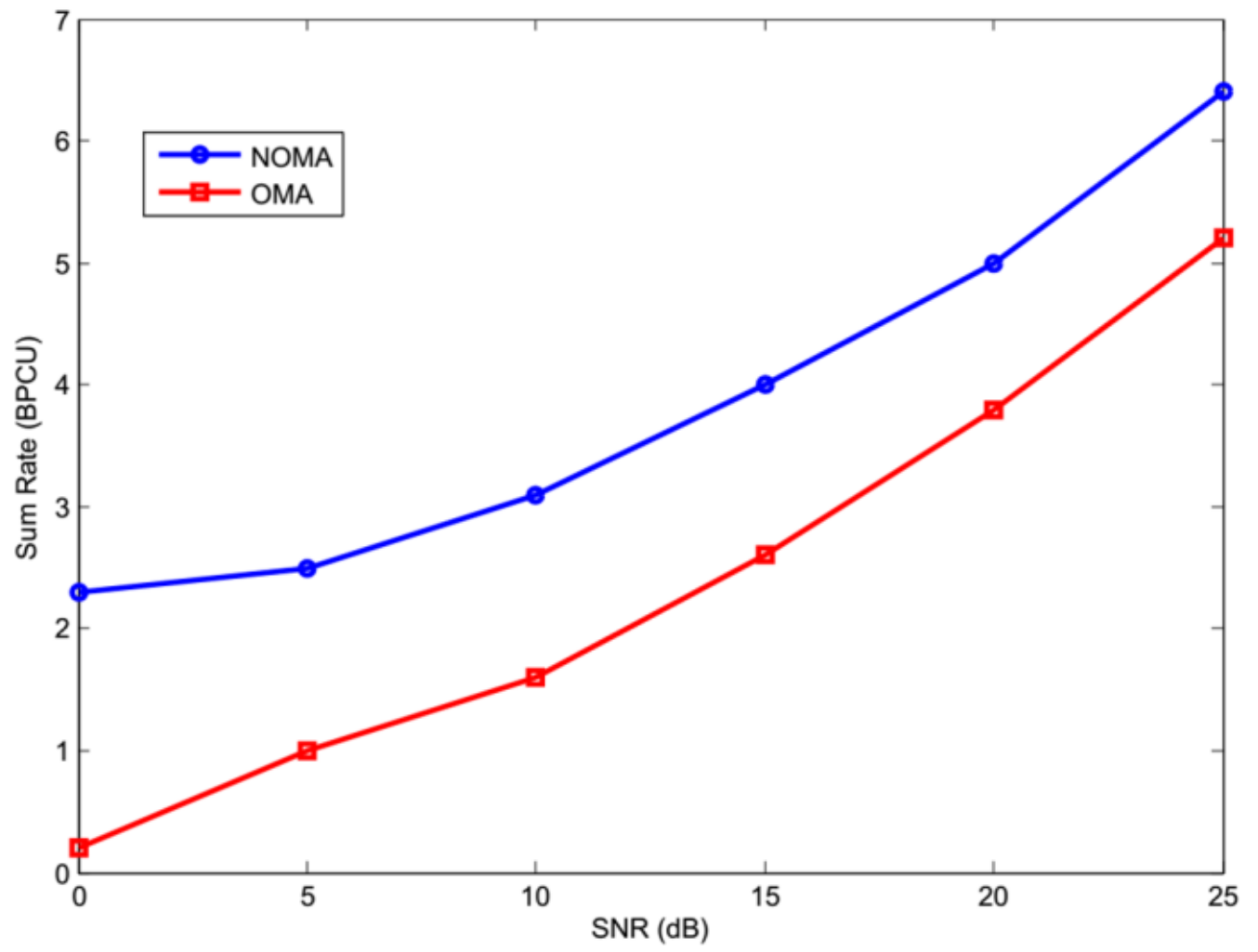

Fig. 12. Sum rate performance of NOMA in 5G with random users in a cell.

Fig. 15 compares the sum rate performance of CSC-based NOMA (discussed in Section III. E) with that of non-CSC-based NOMA under symmetric channel conditions with a path loss exponent of 3. Note that non-CSC–based NOMA considers only one BS (either one) to employ SC to serve a pair of cell-edge and nearby users simultaneously. It is observed that the sum rate of a CSC-based system exponentially increases with SNR, and is higher than a non-CSC–based system.

Fig. 16 compares the performance of user pairing with proportional fairness–based power allocation (discussed in Section III. G) to that of the TTPA scheme [59] in terms of percentage gain with respect to OFDMA. A better performance gain is obtained across the entire range of numbers of users per cell when user pairing is formed through PF-based scheduling. This is because PF-based user pairing effectively preserves user diversity by its power allocation metric. Also, removing invalid user pairs during prerequisite checking perhaps partly contributes to this performance enhancement, since it ensures distinct channel variations among candidate users, to some extent.

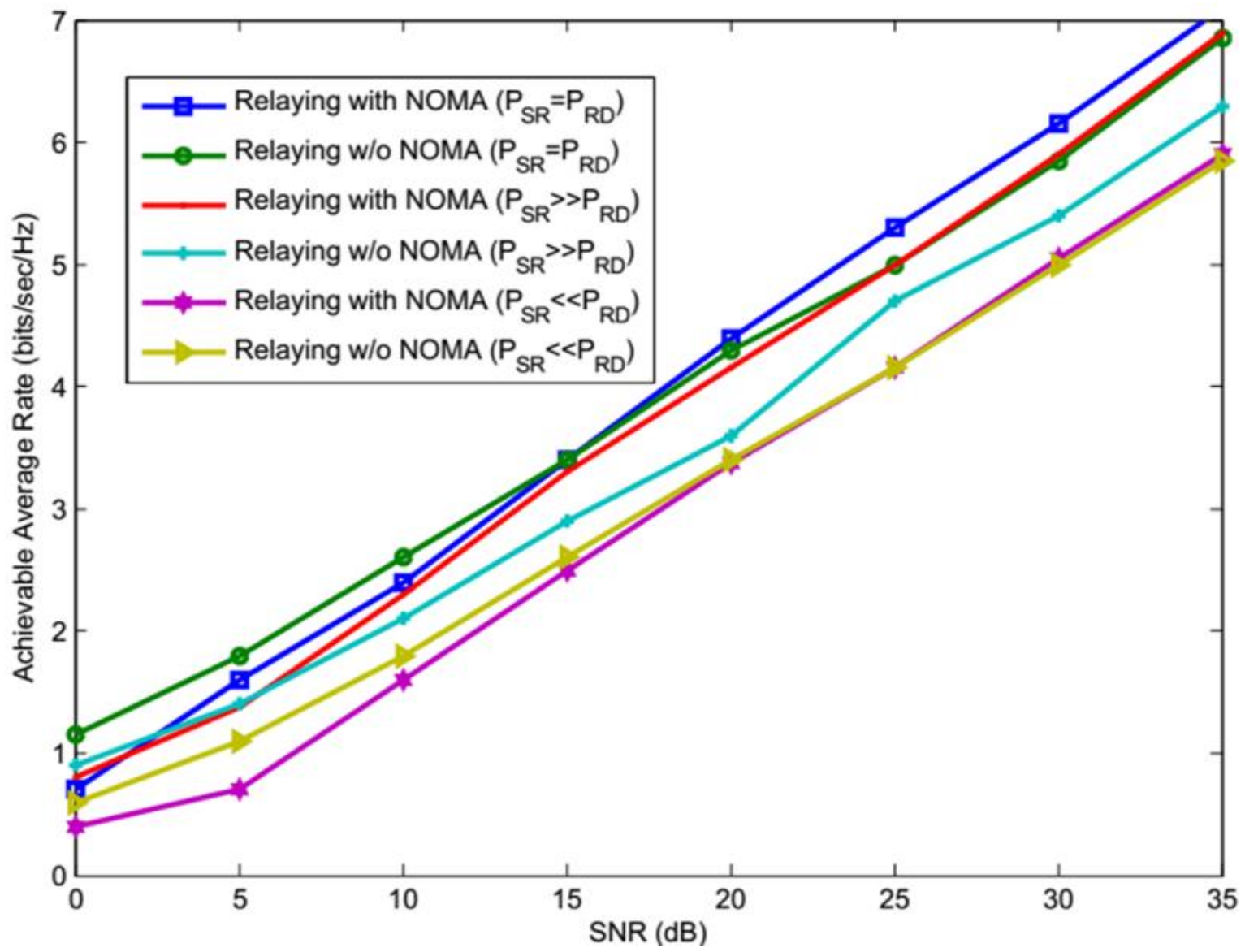

Fig. 13. Achievable rate performance of NOMA with relay.

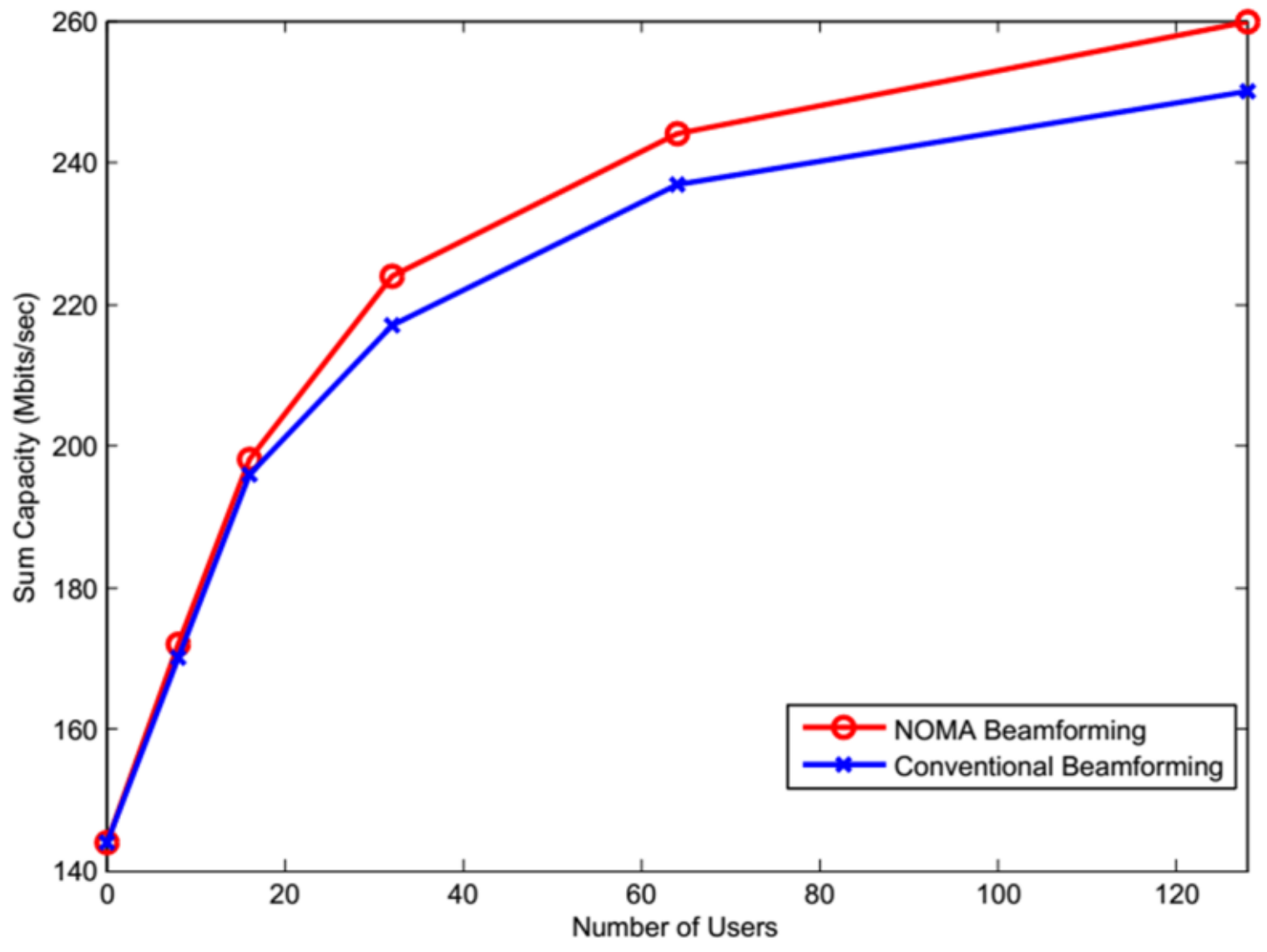

Fig. 14. Sum capacity of NOMA beamforming.

Fig. 17 shows the achievable capacity regions of NOMA and OFDMA-based VLC systems (discussed in Section III. I) with the following numerical parameters: number of LED chips $L_o = 100$ with responsivity $\gamma_o = 0.25$ A/W, and system BW $B = 20$ MHz. In NOMA, the capacity region has been obtained by varying the power coefficients. And in OFDMA, the rate pair was found by varying the BW assigned to each user. As shown, NOMA-based light communications demonstrates higher rate pairs than OFDMA-based VLC. Therefore, it can be concluded that power-domain multiplexing is also promising in light communications.

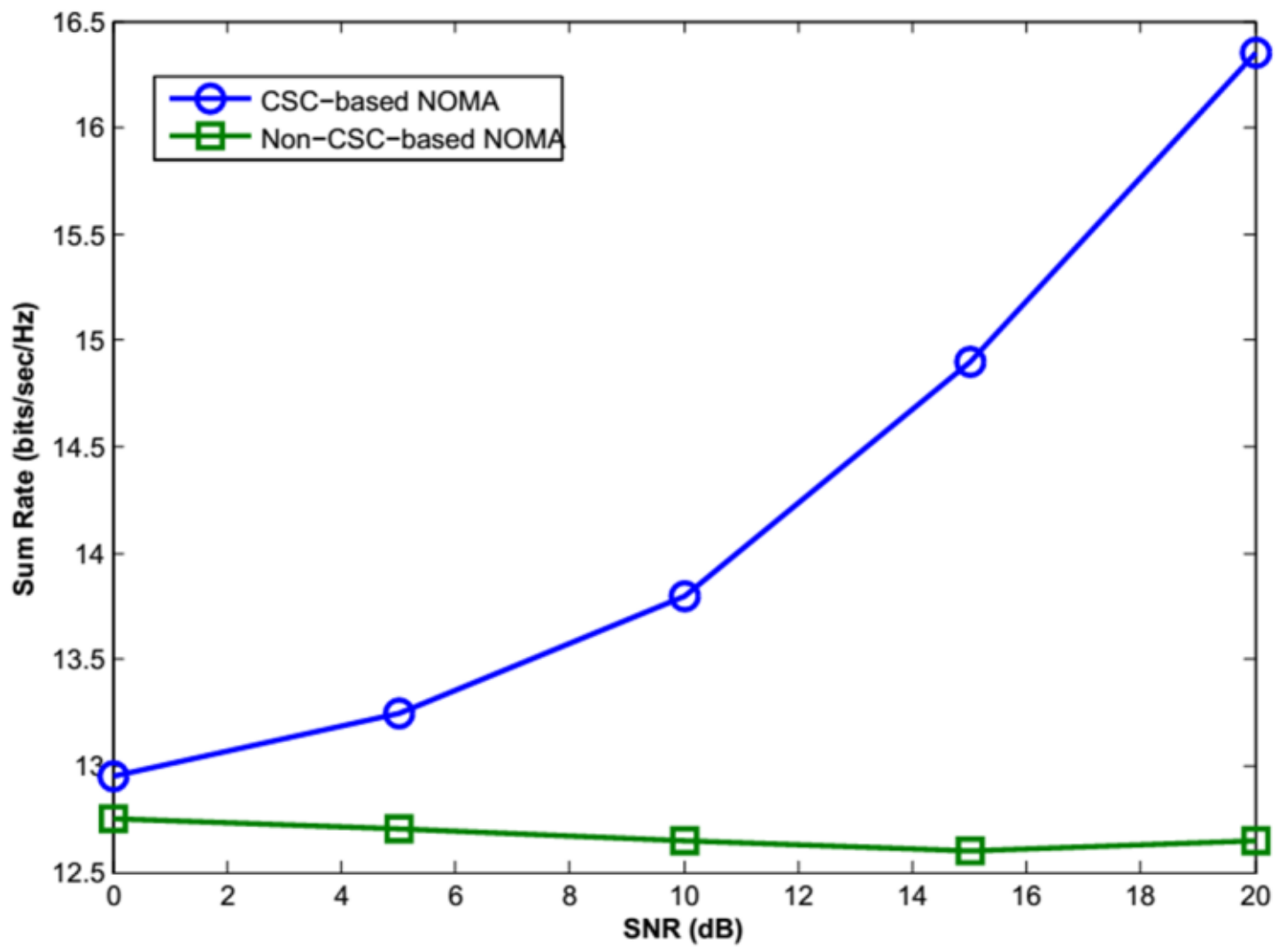

Fig. 15. Sum capacity of CSC-based NOMA.

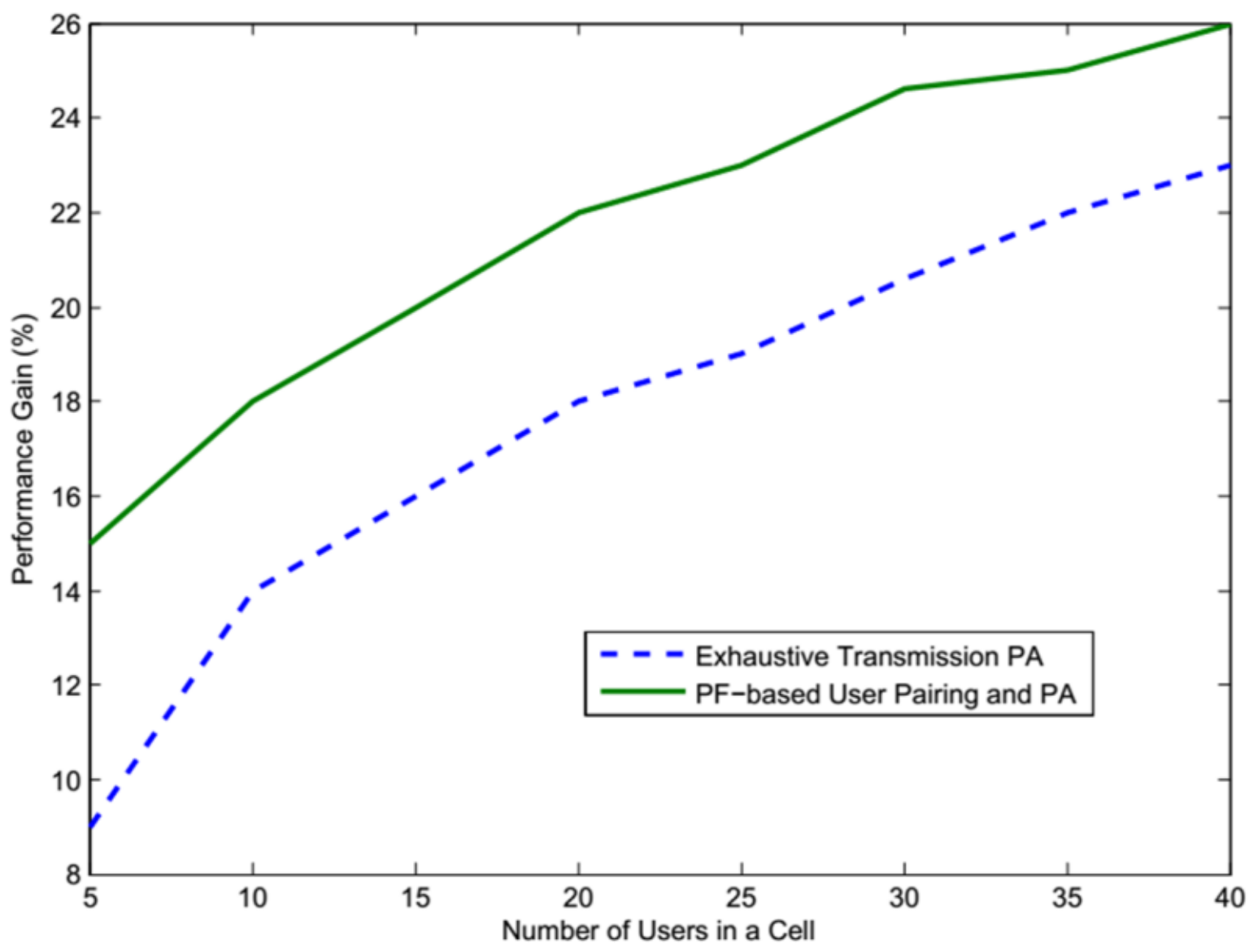

Fig. 16. Performance gain in PF-based user pairing.

### *C. Spectral Efficiency (SE) and Energy Efficiency (EE)*

Suppose the power gains of a $4 \times 2$ downlink channel in the network NOMA scenario described in Fig. 8 are $[-77, -117; -97, -107; -77, -117; -97, -107]$ dB. The SE of this network NOMA scheme is presented in Fig. 18. The baseline to compare the performance of network

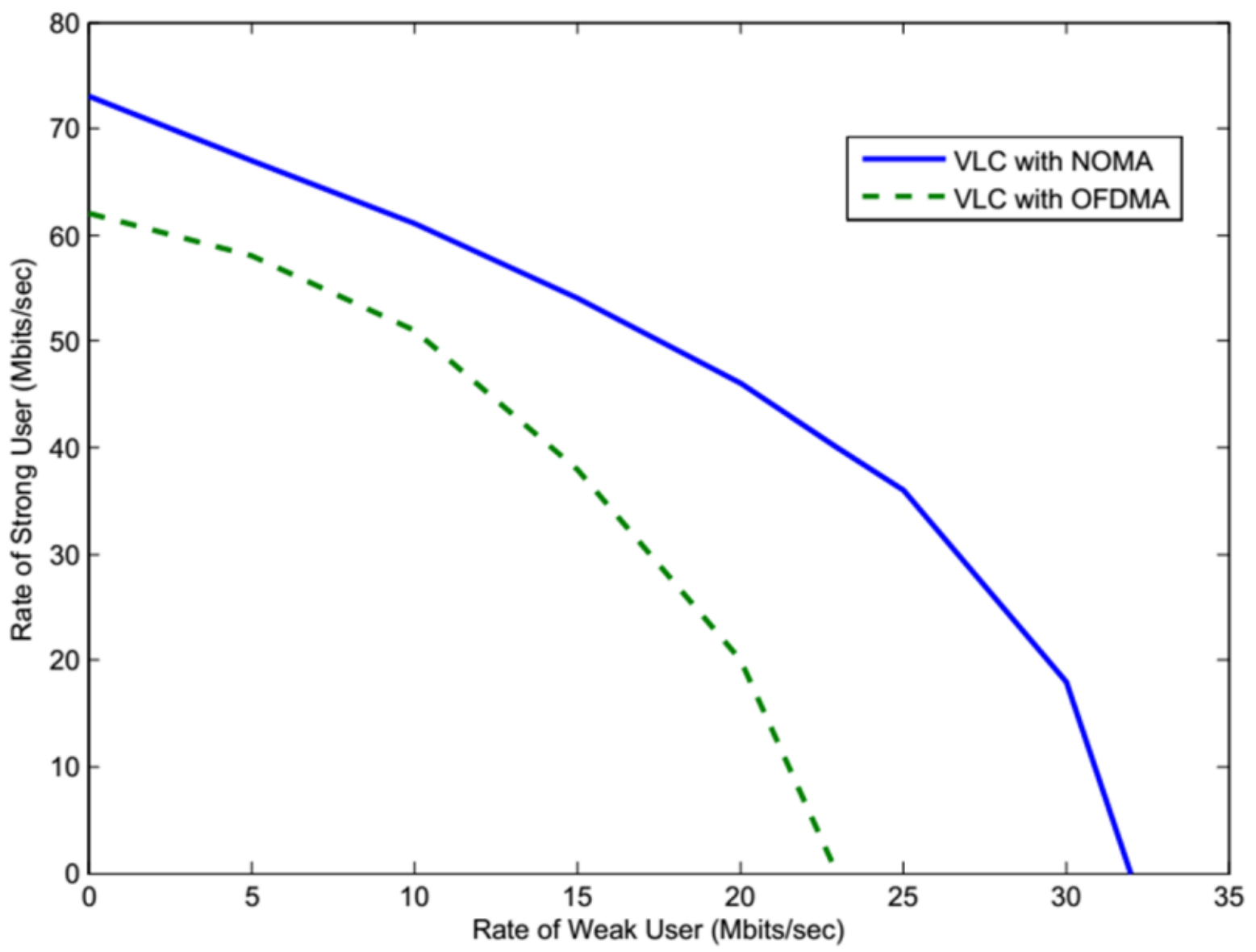

Fig. 17. Capacity regions of light communications with NOMA.

NOMA in terms of total SE is the performance of the entire network based on single-cell NOMA, where base stations do not cooperate with each other. This figure shows that network NOMA outperforms conventional NOMA. The reason is that severe inter-cell interference (mainly between cell-edge users) causes performance degradation in conventional single-cell NOMA, since there is no inter–base-station cooperation. However, the precoder for network NOMA works in such a way that base stations transmit jointly to cell-edge users, and the cell-center users first detect and subtract the signals of cell-edge users to mitigate the mutual interference.

In Fig. 19, the EE-SE trade-off (discussed in Section III. H) curve of a single-user MIMO NOMA scheme is compared with that of single-user MIMO TDMA as an OMA approach to explaining what happens to EE performance when we change spectrum efficiency, and vice-versa. We considered statistical CSI in both cases. The figure demonstrates that NOMA outperforms OMA in terms of better EE-SE trade-off. Both the spectrum efficiency and the EE corresponding to the peak operating point offered by the NOMA scheme are higher values than those achieved by the OMA scheme. The reason is that EE optimization maximizes EE under the constraints of total transmit power and the minimum rate of the weak user with the help of NOMA. The differences in the performance of these two methods become even more significant in high SNR regions.

## V. NOMA Challenges

Many researchers have worked on designing and implementing NOMA techniques and on solving various technological problems associated with those methods. The literature demonstrates that NOMA is compatible with cooperative communications, relaying, and MIMO, and it significantly enhances the performance gains. This paper has found various useful NOMA solutions to address the problems associated with the multi-cell network, wireless link adaptation, EE-SE trade-off, and user pairing. Researchers have introduced a technique where the low data rate and bad quality of services experienced by cell-edge users can easily be improved by adopting NOMA-based coordination. In addition to research concerns in the literature, there are several other challenges and open issues that need to be carefully addressed. This section will try to provide some research directions for researchers interested in investigating NOMA on a larger scale.

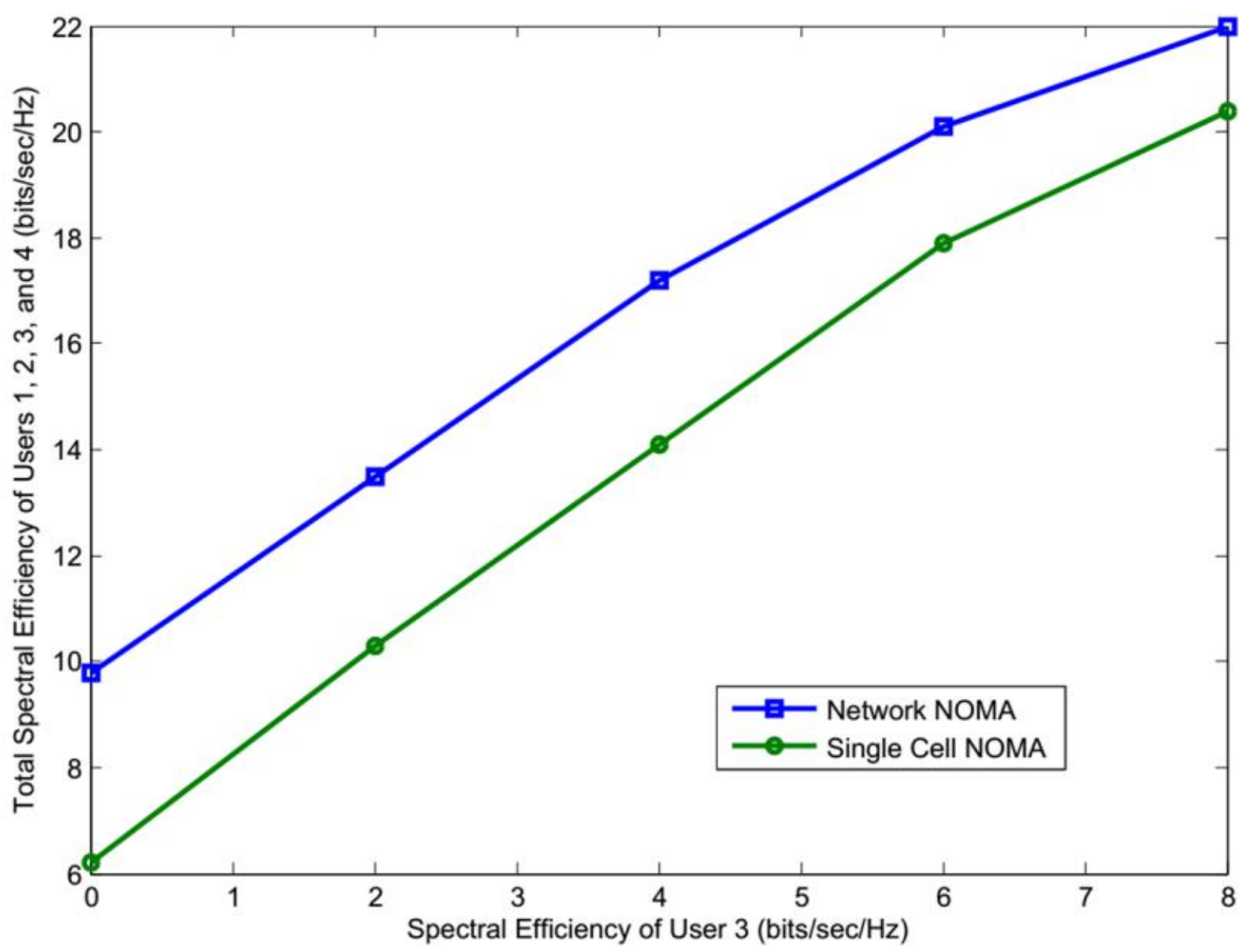

Fig. 18. Spectral efficiency of network NOMA.

### *A. Dynamic User Pairing*

As mentioned earlier in this paper, co-channel interference is strong in NOMA systems, since multiple users share the same time, frequency, and spreading code. As a result, it is difficult to ask all the users in the system to perform NOMA jointly. Alternatively, the users in the system are divided into several groups, where NOMA is applied within each group, and different groups are allocated with orthogonal BW resources. The static case is usually considered, where the $m$th user and the $n$th user, ($|m-n|$ is constant and considerably high) are paired to implement NOMA. Although it is difficult in practice, the design of dynamic user pairing/grouping schemes is necessary to achieve the maximum benefits offered by NOMA. In this regard, the analytical insights obtained by Ding et al. [41] for the case with two selected users can be considered as criteria for the design of distributed methods. It is known that the UPPA scheme [42] improves performance gains. However, only two users are permitted to be scheduled by NOMA. An encouraging extension of the UPPA scheme would be a power allocation scheme in the multi-user NOMA system where the scheduler allows more than two users. Also, it requires more research in order to apply UPPA schemes and variations [44] to a practical system with a suitable modulation and coding scheme.

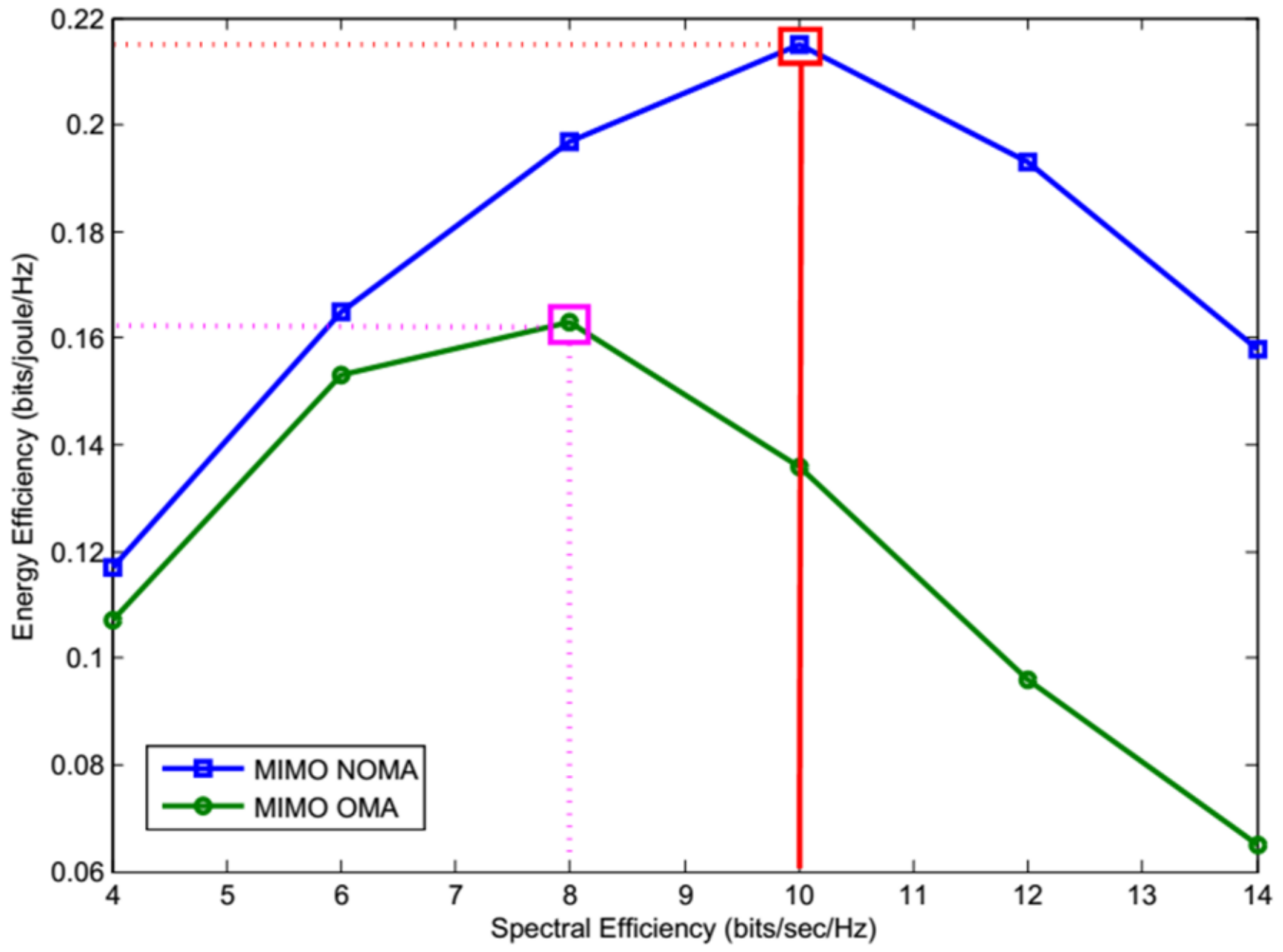

Fig. 19. A trade-off between energy efficiency and spectral efficiency.

### B. Impact of Transmission Distortion

The transmission of source information, such as voice and video over communications channels is generally considered lossy. The transmitted data always experience distortion while they propagate to the receiver. To deal with this lossy transmission, considerable theoretical attention in assessing source fidelity over fading channels has been paid to date. Different source coding and channel coding have been framed to minimize the end-to-end distortion. However, source coding diversity and channel coding diversity provide conflicting situations among the preferred amounts of distortion, cost, and complexity. Choudhury and Gibson compared the source distortion for two definitions of channel capacity, namely, ergodic capacity and outage capacity [60]. Both information capacity and distortion depend on outage probability. It is evident that outage probability that maximizes outage rate may not provide the minimum expected distortion. An investigation can be carried out to optimize the outage probability for which a NOMA scheme can provide the maximum outage rate with acceptable distortion.

### C. Impact of Interference

Although interference analysis is a generic term in wireless communications, here this survey focuses on cooperative NOMA [23], which utilizes Bluetooth-like short-range communications in the cooperative phase. However, the uses of BT radio in cellular communications faces an extreme interference scenario from existing wireless personal area network (WPAN) operations. BT interference decreases the coverage and throughput, causes intermittent or complete loss of connectivity, and results in difficult pairing during the user's discovery phase. In fact, interference in the deployed environment, payload size, and distance between cooperative users all affect the deployment of channel allocation. Also, a self-organizing scatternet to manage BT nodes needs to be reformulated to make it functional with NOMA, since the users in NOMA are paired according to CSI. In addition, a robust scatternet should offer valid routes between nodes with high probability, even though users' mobility causes complete loss of some of the wireless links. Furthermore, due to the mobility of users, interference becomes dynamic. Therefore, performance analysis of a cooperative NOMA scheme in this dynamic interfering environment will be an interesting study.

### D. Resource Allocation

In order to accommodate a diverse set of traffic requirements, 5G systems should be capable of supporting high data rates at very low latency and in reliable ways. However, this is a very difficult task, since resources are limited. So, resource management has to assist with effective utilization. Wireless resource management is a series of processes required to determine the timing and amount of related resources to be allocated to each user [61]. It also depends on the type of resources. According to the Shannon's information capacity theorem, BW is one of these wireless resources. As a part of the effective management of the system BW in a communications system, the total BW is first divided into several chunks. Each chunk is then assigned to a particular user or a group of users, as in the case of NOMA. Also, the number of packets for each user varies over time. Therefore, user-pairing and optimum power allocation among users in NOMA requires a sophisticated algorithm to provide the best performance with the usages of minimum resources. Resource allocation in NOMA can also be explored from a mathematical optimization theory point of view. For example, Lei et al. considered joint power and channel allocation for NOMA in 5G networks in [62]. They take user power control and SIC implementation into account to solve the power and channel allocation problem. Power allocation in NOMA-based cognitive radio networks [63] is also an unexplored area of research.

### *E. NOMA with Multiple Antenna*

The larger the rank of the MIMO channel matrix, the larger the number of de-correlated channels obtained, and thus, better system performance occurs. Therefore, the channel matrix rank plays a key role in evaluating MIMO NOMA. However, existing works in MIMO NOMA consider full-rank channel matrices in order to investigate system performance. Under this constraint, for example, the analytical results developed by Ding et al. [36] provide the upper bounds of outage probability and capacity. It is therefore now essential to study the design of MIMO NOMA with consideration of rank-deficient channel matrices. Rank adjustment for NOMA can be adopted in order to improve MIMO NOMA performance [21]. A joint optimization of rank and transmission power assignment can be investigated to further enhance performance. It is also interesting to focus on existing MIMO NOMA techniques combined with other promising wireless techniques, for example, orthogonal frequency and code division multiplexing (OFCDM). Most of the MIMO NOMA algorithms (for example, the beamforming algorithm proposed by Kim et al. [32]) visited in this paper have very high computational complexity. Therefore, there is an urgent need for research activities into complexity reduction.

### *F. Heterogeneous Networks*

A heterogeneous network (HetNet) is a wireless network consisting of nodes with diverse transmission powers and coverage sizes. The HetNet has potential enough for next-generation wireless networks in terms of capacity and coverage with reduced energy consumption. The infrastructure featuring a high-density deployment of low-power nodes can also significantly increase EE, compared to low-density deployment of fewer high-power nodes. There are several research works on HetNets, for example, node cooperation, optimal load balancing, and enhanced inter-cell interference coordination [64]. A system framework of a cooperative HetNet for 5G was recently studied in [65] with the aims of both spectrum efficiency and EE. Since the objective of NOMA coincides with that of a HetNet, the specific utilization of NOMA in a particular HetNet can offer extended benefits. Also, the non-uniform spatial distribution of mobile users will affect the performance of NOMA. Therefore, investigation of outage performance, ergodic capacity, and user fairness in NOMA schemes with spatial user distribution can be worthy work.

*G. Outage Probability Analysis*

Outage analysis is fundamental to understanding the performance of any wireless system. As a matter of fact, the achievable capacity depends upon the outage behavior of users. Various researchers have investigated the outage probability of the basic NOMA scheme, in general. For example, compared to OMA, less outage occurs in NOMA for users randomly deployed in a cell [19]. That analysis considered the impact of path loss. Also, it was observed that a NOMA-BF system improves the sum capacity, compared to the conventional multi-user BF system [25]. However, when NOMA comes with beamforming, the outage probability of users will be changed. On that, outage performance analysis of NOMA-BF can be investigated. Similarly, network NOMA [40] primarily focuses on inter-cell interference mitigation by designing a precoder. But it needs explicit analysis to understand the outage behaviors of cell-edge users. Other NOMA works, for example, NOMA with VLC [48] and NOMA with coding [52, 53], can be further investigated to analyze outage performance.

*H. Practical Channel Model*

To support the ever-growing amount of consumer data, next-generation wireless networks require not only an efficient radio access technique but also spectrum availability. For the time being, it is obvious that 5G will use spectrum allocations in the unused millimeter wave (mmW) frequency bands. Also, the backbone networks for 5G are expected to move from copper and fiber to mmW wireless connections, allowing rapid deployment and mesh-like connectivity. The mmW frequencies between 30 and 300 GHz are a new frontier for cellular networks that offers a huge amount of BW. Understanding the challenges of mmW cellular communications, in general, and channel behavior in particular, is therefore extremely important, and is a fundamental requirement to developing 5G mobile systems, as well as backhaul techniques [66]. The existing studies on NOMA assume the wireless links between transmitter and receiver exhibit a Rayleigh fading channel with AWGN. A more realistic analysis would be revealed if the measured path loss and delay spread values [67] could be considered to reflect the radio channel in mmW band.

*I. Uniform Fairness*

In mmW cellular, at a distance greater than 175 m, most locations experience a signal outage [66]. Since outage is highly dependent on the environment, actual outage may be more significant if there are more local obstacles. Deriving a NOMA scheme that provides users (especially located at a distance greater than 150 m up to the cell boundary, in the case of mmW cellular) uniform outage experiences would be excellent work.

*J. NOMA with Antenna Selection*

The simultaneous use of multiple antennas on the transmitter side requires a corresponding number of parallel radio frequency (RF) chains at the front end. As such, it increases system complexity, power consumption, and cost with an increase in the number of antennas. To overcome such issues, transmit antenna selection (TAS) is often preferred to make the transmitter structure simple. An effective TAS technique opportunistically selects the best antenna out of multiple antennas and uses a single RF chain for transmission. This eventually reduces system complexity, power consumption, and cost, as well as size, at the expense of acceptable performance loss. In this respect, research should be conducted into finding a novel TAS-NOMA scheme for downlink communications from a base station equipped with multiple antennas to multiple users, each equipped with a single antenna. The aim of the scheme should ultimately improve the sum rate, considering the targeted user rate is allocated opportunistically based on channel conditions.

### *K. Carrier Aggregation*

In order to increase the BW and, thereby increase the data rate of its users, LTE-Advanced utilizes the concept of carrier aggregation (CA) [68]. The CA concept is that users are allocated aggregated resources consisting of two or more component carriers (CCs). A CC is nothing but each aggregated carrier. The arrangement of aggregation would be either contiguous allocation, where CCs are set adjacent to each other, or non-contiguous allocation, where there is a gap in between. It is possible to integrate CA with NOMA in order to take the advantages offered by both. To do so, however, the user pairing will be different from that of basic NOMA. It is well known that a NOMA user is paired with another NOMA user in a basic two-user NOMA scheme; each carrier is usually assigned to two users based on channel conditions. The user paring in CA-enabled NOMA can be explained as follows. A CC (say, $f_{AN1}$) assigned to a NOMA user ($U_{AN1}$) will also be a CC of another NOMA user ($U_{AN2}$). The situation is that the user $U_{AN1}$ is paired with $U_{AN2}$ with respect to the CC $f_{AN1}$. At the same time, it is also possible that the user $U_{AN1}$ is paired with $U_{AN3}$ with respect to the CC $f_{AN2}$. On that, a NOMA user might be pared with multiple different users at the same time based on the number of CCs if CA is integrated with basic NOMA. To the best of authors' knowledge, answering the question of what CA type is appropriate for NOMA solutions represents an open issue, and thus, investigation of different types of CA in NOMA is an interesting research direction.

### *L. Other Challenges*

There are also some other challenges that need to be addressed before NOMA becomes a part of 5G in the future. In a downlink scenario, for example, the transmitter allocates the power to the users based on their respective CSI. Therefore, a proper mechanism for CSI feedback, a suitable channel estimation scheme with proper reference signal design, is important for achieving robust performance. In addition, the benefit of NOMA is proven only under the ideal condition of perfect acquisition of CSI data at the transmitter side. One possible solution might be the use of a limited feedback channel to acquire CSI [69]. However, it requires additional BW to convey various channel quality indicators. From this perspective, proper user selection and power allocation schemes are necessary when the transmitter has to work under imperfect CSI and/or with limited feedback. In multicarrier communications, the peak-to-average-power ratio (PAPR) can cause the transmitter's power amplifier (PA) to run within a non-linear operating region. This causes significant signal distortion in PA output. The effect of PAPR is thus critical to determining what techniques to use to achieve the best NOMA performance. The concept of relaying [29] to extend cell coverage can be applied in non-orthogonal coordinated transmissions between multiple small cells and a macro cell in order to achieve more capacity gain. A hybrid scheme proposed by Kalokidou et al. [70] combines the principles of topological interference management and NOMA schemes to achieve better performance in terms of sum rate. Further work on the MIMO form of this hybrid method can be investigated. However, the challenge, especially for a dense network, is how to determine a fair power allocation method. To adopt NOMA in 5G, NOMA should also be made robust in terms of system scalability, since 5G must support heterogeneous traffic and diverse radio environments. To date, most of the work has been dedicated to verifying the performance of NOMA in theory, but not in practice. Although Xiong et al. introduced a software-defined radio (SDR) NOMA as a prototype [71], more over-the-air experiments are required for the purpose of demonstrating NOMA's potential in 5G mobile networks.

## VI. Implementation Issues

NOMA implementation includes designing and operating the system. A lot of what will be attempted at this stage will rely upon a particular application and environment. In this section, this survey tries to give some generic points, including some on computational complexity and error propagation that can be critical to assessing NOMA performance.

### *A. Decoding Complexity*

Signal decoding by using SIC requires additional implementation complexity compared to orthogonal schemes, since the receiver has to decode other users' information prior to decoding its own information [3]. Also, this complexity increases as the number of users in the cell of interest increases. However, users can be clustered into a number of groups, where each cluster contains a small number of users with bad channels. SC/SIC can then be executed within each group. This group-wise SC and SIC operation basically provides a tradeoff between performance gain and implementation complexity.

### *B. Error Propagation*

It is intuitive that once an error occurs in SIC, all other user information will likely be decoded erroneously. However, the effect of error propagation can be compensated by using a stronger code (e.g., increasing the block length) when the number of users is reasonably small [3]. In case of degradation in the performance for some users, nonlinear detection techniques can also be considered to suppress the error propagation. Based on computer simulations, authors in [2] showed that the error propagation can have a marginal impact on the NOMA performance. The reason is that a user with bad channel gain is assigned to another user with good channel gain during NOMA scheduling. This finding has been reported under the conditions of a worst-case model which assumes that the decoding of the strong user at the second stage is always unsuccessful whenever the decoding of the weak user is unsuccessful at the first stage of the strong user receiver. Although there exist works that analytically study the SIC error propagation in basic MIMO systems [72, 73], there is no prominent research that provides a mathematical understanding of the effect of imperfect SIC on NOMA schemes. Therefore, a mathematical analysis of the impact of imperfect SIC on NOMA performances represents an interesting research direction.

### *C. Quantization Error*

When the received signal strengths of users are disparate, the analog-to-digital (A/D) converter needs to support a very large full-scale input voltage range, and requires high resolution to accurately quantize a weak signal, since the more levels the ADC uses for quantization, the lower its quantization noise power. However, there is a limitation placed on arbitrarily high–resolution ADC due to its cost, conversation time, and hardware complexity. This constraint eventually leads to a trade-off between the quantization error and SIC gain.

### *D. Power Allocation Complexity*

The achievable throughput of a user is affected by the transmit power allocation to that user. This particular power allocation also affects the achievable capacity of other users, since the basis of NOMA is power-domain user multiplexing. To achieve the best throughput performance in NOMA, a brute-force search over the possible user pairs with dynamic power allocation is required. However, this kind of exhaustive search is computationally expensive.

### *E. Residual Timing Offset*

Synchronous transmission has been considered in NOMA research, and this consideration is reasonable for the downlink scenario, because the BS controls transmission for all users. However, perfect synchronization among NOMA users is impractical on uplink, since users are spatially distributed, and the mobile communications channel is usually dynamic in nature. In asynchronous communications, OFDM symbols from the superposition-coded users are time-misaligned. Thus, NOMA users' performance substantially depends on the relative time offset between interfering users [74]. On that, it is important to further investigate NOMA in asynchronous communications. In practice, the asynchronous scheme requires information on multiple symbols for detection and interference cancellation. Otherwise, performance degrades when complete information on interfering signals is unknown.

### *F. NOMA or IDMA*

Both NOMA and interleave division multiple access (IDMA) are prospective multiple access schemes for equal rate transmission without link adaptation. The question may arise as to which scheme should be adopted in practical system implementation. Chen et al. concluded that IDMA is able to deal with the user power–balanced scenario and still offer robust performance with higher computational complexity [75]. And NOMA is effective in the power-imbalanced scenario, smartly performing the trade-off between performance and complexity.

### G. Signaling and Processing Overhead

There are several sources of additional signaling and processing overhead in NOMA compared to its orthogonal counterparts. For example, to collect the CSI from different receivers and to inform the receivers of the SIC order, some time slots need to lapse. This causes data rate degradation in NOMA. Also, with dynamic power allocation, and encoding and decoding for SC and SIC, NOMA signal processing requires additional energy overhead.

### H. Limited Number of User Pairs

NOMA exploits power-domain multiplexing to offer its benefits. More specifically, about an 8 dB difference in propagation loss is required to pair a cell-edge user with a cell-center user [76]. Therefore, the possible number of pairs of users in a typical NOMA scheme is limited, which eventually reduces the capacity gain of NOMA. Finding the appropriate signal detection and decoding strategy that increases the number of pairs of users is an important issue.

### I. NOMA Deploying Environment

Since the power domain is exploited, and channel gains are naturally adequately large in macrocells, researchers commonly consider macrocell deployment. In addition to macrocells, small cells are also becoming important and are being studied for future 5G networks. Fortunately, the NOMA gain is still achievable, even in a small cell where NOMA offers a higher performance gain compared to OMA [77]. It is also important to note that NOMA gain in terms of cell throughput in small cells is even larger than in macrocells. Therefore, NOMA can be equally deployed in both macrocell and small-cell environments.

### J. Standardization Status

There exist some standardization efforts on NOMA technologies. Below is the summary of the current status of ongoing standardization of NOMA. The 3rd Generation Partnership Project (3GPP) LTE Release 13 approved the study item (SI) of downlink multiuser superposition transmission (MUST) [78]. The primary objective of MUST was to identify promising enhancements of downlink multiuser transmission schemes within one cell. To achieve this objective, the SI focused on the evaluation of system level gain and complexity-performance trade-off under practical deployment scenarios and traffic models [79, 80, 81, 82, 83, 84]. The study concludes that NOMA can increase system capacity and improve user experiences. A new work item (WI) of downlink multiuser superposition transmission for LTE has recently been approved by 3GPP LTE Release 14 [85]. The core objective of this WI is to identify necessary techniques to enable LTE to support downlink intra-cell multiuser superposition transmission for the physical downlink shared channel.

## VII. Concluding Remarks

This paper provides a comprehensive overview of the present and emerging power-domain

SC-based NOMA research into 5G, and discusses NOMA performance with numerical results. It is clear that NOMA is a candidate multiple access technology for next-generation radio access. Its diversity gain originates from the power domain of the signals to be transmitted in a superposed fashion. Many research results have been found in favor of NOMA in terms of outage probability, achievable capacity, weak users' rate guarantees, and cell-edge user experiences. In addition to perfect SC at the transmitter and error-free SIC at the receiver, optimum power allocation, QoS-oriented user fairness, appropriate user pairing, and good link adaptation are also required to obtain the maximum benefits offered by NOMA. In addition, this paper discusses how NOMA works with various standard wireless technologies, including cooperative communications and MIMO. For a deeper understanding of NOMA, this paper provides a discussion on how inter-cell interference in a network can be mitigated, and explains how a trade-off between energy efficiency and bandwidth efficiency can be achieved. The discussions of several important issues, such as dynamic user pairing, distortion analysis, interference analysis, resource allocation, heterogeneous networks, carrier aggregation, and transmit antenna selection, are expected to facilitate, and provide a basis for, further research on NOMA in 5G. This paper offers a general view of some implementation issues, including computational complexity, error propagation, deployment environments, and standardization status. In sum, the results of this survey are expected to be useful to researchers working in the area of wireless communications and NOMA.

## ACKNOWLEDGEMENT

The authors would like to thank the anonymous reviewers for their valuable comments and suggestions to improve the quality of the paper.